\documentclass[12pt]{iopart}
\usepackage{graphicx}
\usepackage{iopams}
\usepackage{amsthm}
\usepackage{epsfig}
\usepackage[usenames]{color}
\usepackage{mdwlist}
\usepackage{algorithmic}

\newtheorem*{pf}{Proof}
\newtheorem{thm}{Theorem}

\newtheorem{lm}{Lemma}

\newtheorem{crl}{Corollary}
\newtheorem{rmk}{Remark}

\newtheorem{example}{Example}

\newcommand{\scprod}[2]{\left<#1|#2\right>}
\newcommand{\ket}[1]{\vert#1\rangle}
\newcommand{\bra}[1]{\langle#1\vert}

\newcommand{\RR}{{\mathbb R}}

\newcommand{\NN}{{\mathbb N}}
\newcommand{\CC}{{\mathbb C}}
\newcommand{\SScirc}{\mathbb{S}^1}

\newcommand{\SSS}{{\mathcal{S}}}
\newcommand{\abs}[1]{\left|#1\right|}
\newcommand{\norm}[1]{||#1||}

\newcommand{\dotex}{\frac{d}{dt}}

\newcommand{\proj}[1]{\vert#1\rangle\langle#1\vert}
\newcommand{\sigmax}[2]{\vert#1\rangle\langle#2\vert+\vert#2\rangle\langle#1\vert}
\newcommand{\minou}{\mbox{-}}

\begin{document}

\title{Adiabatic passage and ensemble control of quantum systems}

\author{Z Leghtas$^1$, A Sarlette$^{2,3}$\footnote{Corresponding author A Sarlette, Tel:+32~43662972, Fax:+32~43662989.} and P Rouchon $^3$}

\address{$^1$ INRIA Paris-Rocquencourt Domaine de Voluceau, BP105 78153 Le Chesnay Cedex, France. Email: zaki.leghtas@inria.fr}
\address{$^2$ Department of Electrical Engineering and Computer Science, University of Li\`ege, 4000 Li\`ege, Belgium. Email: alain.sarlette@ulg.ac.be}
\address{$^3$ Mines-ParisTech, Centre Automatique et Syst\`emes, 60, boulevard Saint-Michel 75272 Paris Cedex, France. Email: pierre.rouchon@mines-paristech.fr}

\begin{abstract}
This paper considers population transfer between eigenstates of a finite quantum ladder controlled by a classical electric field. Using an appropriate change of variables, we show that this setting can be set in the framework of adiabatic passage, which is known to facilitate ensemble control of quantum systems. Building on this insight, we present a mathematical proof  of robustness for a control protocol -- chirped pulse -- practiced by experimentalists to drive an ensemble of quantum systems from the ground state to the most excited state. We then propose new adiabatic control protocols using a single chirped and amplitude shaped pulse, to robustly perform any permutation of eigenstate populations, on an ensemble of systems with badly known coupling strengths. Such adiabatic control protocols are illustrated by simulations achieving all 24 permutations for a $4$-level ladder.
\end{abstract}


\section{Introduction}

Population transfer from eigenstate $k$ to eigenstate $l$ of a quantum system refers to finding a control input such that the projection of final system state on eigenstate $l$ of the free Hamiltonian has the same norm as the projection of initial system state on eigenstate $k$. Applications of population transfer range from population inversion \cite{Hare93}, where $k$ and $l$ are lowest and highest energy eigenstates, to quantum information processing \cite{NielsenChuang,Averin98,Aharonov07}, where logic gates would (selectively) permute the populations of several eigenstates. In many applications, including those mentioned, relative insensitivity to variations in system parameters is important for robustness issues.

In the present paper, we show how control inputs designed on the basis of adiabatic passage can implement any given permutation of eigenstate populations for a finite anharmonic quantum ladder. The controls we use are chirped pulses
\cite{Chelkowski-al-PRL_1990}
with appropriately modulated amplitudes and exploit the idea of eigenvalue crossing \cite{Yatsenko-Guerin-Jauslin-PRA_2002}. The ladder consists of a free Hamiltonian with approximately equidistant eigenvalues and where the control input couples eigenstates associated to consecutive eigenvalues. A striking robustness feature is that our control fields must only satisfy a set of key properties and achieve population transfer independently of the values of dipole moments coupling consecutive levels of the ladder.
This is a major difference with respect to early non-adiabatic approaches to molecular ladder dissociation using chirped pulses \cite{Chelkowski-al-PRL_1990}.

In this sense, we achieve a specific form of \emph{ensemble control}. Ensemble control in its most general form wants a same input to drive an ensemble of systems, with different values of some parameter $p$, from given $p$-dependent initial state to given $p$-dependent final state \cite[definition 1]{Li-Khaneja-IEEE_2009}. Currently, solutions to this general problem are essentially restricted to two-level systems, achieving approximate ensemble control in finite time and exact ensemble control in infinite time \cite{Li-Khaneja-IEEE_2009,Li-Khaneja-PRA_2006,Beauchard-Coron-Rouchon_2010}. They rely on
accurate knowledge of laser-system coupling strengths
and accurately tailored inputs, involving e.g.~exact instantaneous ``$\pi$-amplitude-impulses''.
In our setting, system parameters need not be exactly known and the input must only satisfy a few key properties. In turn, regarding initial-to-final-state transformations, we are limited to population permutations (with arbitrary relative phases between components of different eigenstates) that are constant as a function of system parameters.
Driving an ensemble of 2-level systems from a common initial to a common final state has also been much studied in the NMR context, e.g.~with geometric methods \cite{Li-Khaneja-PRA_2006}.

\emph{Adiabatic passage} is a control strategy that builds on the adiabatic evolution property: A system state initially close to an eigenstate of a time-varying Hamiltonian $H(t)$ approximately follows the time-varying eigenstate of $H(t)$ if it varies slowly enough; the slower $H(t)$ varies, the better the adiabatic approximation. A thorough formal study of adiabatic evolution can be found in \cite{Avron87,Avron99,Teufel_2003}, on which we build the proofs of our results. Adiabatic evolution has been standard since the early days of quantum mechanics \cite{Messiah58}, e.g.~when interpreting system evolution in terms of ``avoided eigenvalue crossings''.
In a ladder control context, population inversion in two-level systems by a ``chirped'' pulse
--- where frequencies of a Gaussian laser pulse are spread out in time --- is known by experimentalists and theoretically explained in the adiabatic framework \cite{Allen-Eberly_1987}. This is the most basic case of our control, section~\ref{sec:0toN} with $N=2$. Many experimentalists have then focused on \emph{multiple-laser techniques}, individually addressing pairwise couplings in an $N$-level system; this includes stimulated Raman adiabatic passage (STIRAP), see e.g.~\cite{Oreg84,ShapiroTmultilaser,Vitanov01}. For $N$-level ladder systems specifically, the possibility of population transfer from the lowest to the highest energy eigenstates with a single \emph{chirped} laser pulse has been recognized and exploited in ``adiabatic rapid passage'' experiments \cite{Hulet83,Hare93,Maas-al-Noordam-PRA_1999,Vitanov01}.
An analysis of $N$-level adiabatic molecular dissociation with chirped pulses is given in \cite{Guerin-PRA_1997} based on the Floquet representation.
In the present paper we provide a simple mathematical proof of population inversion with avoided crossings (gap condition) based on Favard's Theorem \cite{Favard-AcadScParis_1935} and on the roots of orthogonal polynomials \cite{Szego-AMS_1967},
and extend the framework by adding amplitude control to perform not only population inversion but all different permutations of free Hamiltonian eigenstates.

The paper is organized as follows.
Section~\ref{sec:0toN} gives the formal statement and section \ref{sec:proofs} the proof for $N$-level population inversion with ``adiabatic rapid passage'', actually proving how initial population of level $k$ is finally transferred to level $N-k-1$ in adiabatic approximation. The key point for using adiabatic passage is a change of frame that depends on time-varying control input phase; it is detailed in section~\ref{sec:problem setting} after formal description of the ladder system. The proof then applies the standard ``adiabatic theorem with spectral gap condition'', where time-varying eigenvalues are shown to remain separated for all times.
The inversion is insensitive to exact energy values of the individual levels in the ladder.
Section~\ref{sec:ltop} proposes adiabatic control inputs to transfer population between two arbitrary eigenstates. It requires the control field to vanish at specific times which depend on (some) energy levels of the \emph{anharmonic} ladder, such that we select a pair of time-varying eigenvalues to cross. System evolution is then ruled by the ``adiabatic theorem without spectral gap condition''. A complementary study
of system behavior in the neighborhood of two crossing eigenvalues and valid for more general systems than ladder ones, can be found in \cite{BoscainCDC2010}.
We again provide a formal proof of the control's effect and highlight its ensemble/robustness features in section \ref{sec:proofs}.  Section~\ref{sec:permutations} finally shows how any permutation of eigenstate populations can be achieved in this adiabatic passage framework. Each control protocol is illustrated by a simulation at the end of the corresponding section.


\paragraph*{Notation:} We use the Dirac bra-ket notations: $\ket{\psi}\in\CC^N$ denotes a complex vector, $\bra{\psi}=\ket{\psi}^\dag$ is its Hermitian transpose, and $\scprod{.}{.}: \CC^N\times\CC^N\rightarrow\CC : (\ket{\psi_1},\ket{\psi_2}) \rightarrow \scprod{\psi_1}{\psi_2}=\bra{\psi_1} \ket{\psi_2}$ is the Hermitian scalar product. For $z\in\CC$ we note $\Re(z)$ its real part and $z^*$ its conjugate. $\mathcal{H}_N$ is the set of $N\times N$ Hermitian matrices, where $N\in\NN$. We note $I$ the $N\times N$ identity matrix.
For any matrix $A \in \CC^{N \times N}$, we denote its Frobenius (or Hilbert-Schmidt) norm $\Vert A \Vert = \sqrt{\tr{A^\dag \, A}}$ where $\tr{\cdot}$ denotes trace. For $H \in \mathcal{H}_N$, it holds
$\Vert H \Vert = \sqrt{\sum_{i=0}^{N-1}\; \lambda_i^2}$ where $\lambda_0,\ldots,\lambda_{N-1}$ are the (real) eigenvalues of $H$.
For $H\in\mathcal{H}_N$ and $\lambda$ an eigenvalue of $H$, we denote $P_{\lambda} \in\mathcal{H}_N$ the orthogonal projector on the eigenspace of $H$ associated to the eigenvalue $\lambda$. If $H$ has $M$ distinct eigenvalues $\{\lambda_0,..,\lambda_{M-1}\}$, with $M \leq N$, then $H=\sum_{k=0}^{N-1}{\lambda_k P_{\lambda_k}}$ is the spectral decomposition of $H$. If $M=N$, then $H$ is called \emph{non degenerate} and each $P_{\lambda_k}$ is a rank-one projector. When $M < N$ we say that $H$ is degenerate; then some $P_{\lambda_k}$ have rank larger than $1$.\\
$\SScirc$ denotes the unit circle equivalent to $\RR$ modulo $2\pi$. For $J$ an interval of $\RR$, the derivative of a differentiable function $f: J \rightarrow \SScirc$ is a function from $J$ to $\RR$.
For all $n\in\NN$, we denote $\mathcal{C}^n(J,K)$ the set of $n$ times continuously differentiable functions from $J$ to $K$, where $J$ is an interval of $\RR$ and $K$ is an interval of $\RR$ or $\SScirc$. A multi-component function is $n$ times continuously differentiable, e.g.~$H(s) \in \mathcal{C}^n(J,\mathcal{H}_N)$, if all its components belong to $  \mathcal{C}^n(J,K)$.
For $f\in\mathcal{C}^1(J,K \subseteq \RR^n)$, we note $f'(y)\in\mathcal{C}^0(J,\RR^n)$ the value at $y \in J$ of the derivative of $f$.
$\RR_{>0}$ is the set of strictly positive real numbers; we use analog notation with $\geq$, $\leq$ or $<$. $\NN_a^{b}$ is the set of integers from $a$ to $b$, both boundaries included.
When writing $c_0,\ldots,c_{N-1} \in \SSS$ we mean that $c_k$ belongs to the set $\SSS$ for each $k \in \NN_0^{N\minou 1}$. Infimum and supremum of a set are noted $\sup$ and $\inf$ respectively.


\section{Problem setting}\label{sec:problem setting}


\subsection{Standard formulation}\label{ss:standard}

Consider a quantum system with wavefunction $\ket{\psi}\in\CC^N,~\scprod{\psi}{\psi}=1,~N\in \NN$, whose dynamics is governed by the Schr\"{o}dinger equation (with $\hbar=1$)
\begin{equation}
\label{eq:dyn_psi}
i \, \dotex \ket{\psi(t)}=(H_0+u(t)H_1)\, \ket{\psi(t)} \, .
\end{equation}
The Hamiltonians $H_0\in\mathcal{H}_N$ and $H_1\in\mathcal{H}_N$ respectively characterize free and control-induced evolution, $u(t)$ being a real scalar control.
In the present paper, we consider a quantum ladder for which the Hamiltonians, in the eigenbasis $\{\ket{0},\ldots,\ket{N-1}\}$ of $H_0$, take the form
\begin{eqnarray}
H_0&=&\sum_{k=0}^{N-1}{k(\omega_0+\Delta_k)\, \proj{k}}\label{eq:H0}\\
H_1&=&\sum_{k=0}^{N-2}{\mu_{k}\, (\sigmax{k}{k+1})}\label{eq:H1} \; ,
\end{eqnarray}
with $\omega_0\in \RR_{>0}$; $\Delta_0,\ldots,\Delta_{N-1}\in\RR$; and $\mu_0,\ldots,\mu_{N-2} \in \RR_{>0}$. We assume that system \eref{eq:dyn_psi} features two very different orders of magnitude,
\begin{equation}
\label{eq:orders}
\norm{u(t)H_1}\approx\abs{\Delta_k}\ll\omega_0 \;\; \mbox{ for all } k \mbox{ and all } t \, .
\end{equation}

Physically, $H_0$ is the free Hamiltonian of a quantum ladder with mean resonant frequency $\omega_0$ and anharmonicities $\Delta_k$. We call eigenstates $\ket{0},\ldots,\ket{N-1}$ of $H_0$ the \emph{levels} of the ladder. $H_1$ is the dipole moment matrix and models couplings between consecutive eigenstates; it is therefore tridiagonal with zero diagonal elements, and can be taken real positive and symmetric without loss of generality.
Condition \eref{eq:orders} expresses that control amplitude is relatively weak and that the ladder is close to a harmonic one, i.e.~eigenvalues of $H_0$ associated to consecutive eigenstates are close to equidistant. This allows to exploit resonant transitions between all consecutive eigenstates with a control of carrier frequency $\omega_0$.
We consider a typical such control with a small positive parameter $\varepsilon$,
\begin{eqnarray}
\label{eq:control}
u(t) & = & 2\, \Re\left(e^{i\omega_0t}E(t)\right)\; , \;\;\; E(t)=A(\varepsilon t)e^{\frac{i}{\varepsilon}\theta(\varepsilon t)} \\
\label{eq:contineq} & & \mbox{with } \Vert \dotex E(t) \Vert \ll \omega_0 \; ,
\end{eqnarray}
where $A(t)\in \RR$ and $\theta(t)\in \SScirc$ for all $t\in \RR_{\geq 0}$. Parameter $\varepsilon$ governs the rate of variations in the envelope $A(\varepsilon t)$ and frequency $\frac{d}{dt}\frac{1}{\varepsilon}\theta(\varepsilon t)=\theta'(\varepsilon t)$ of $E(t)$; we show in the next sections how taking $\varepsilon$ small allows to apply adiabatic passage properties. The slow but nonzero frequency variation is a key element for our control strategy.
Physically, control fields like \eref{eq:control} are obtained e.g.~by ``shaping'' a single laser pulse \cite{Weiner-ScInst_2000}.

The rotating wave approximation (RWA), standard in quantum systems modeling, consists in writing \eref{eq:dyn_psi} with the change of variable $\; \ket{\phi(t)}= \; \left(\sum_{k=0}^{N-1}
\; e^{ik \omega_0 t} \, \proj{k} \right) \;\ket{\psi(t)} \;$ and neglecting fast oscillating terms, to keep only those that vary at frequencies $\ll \omega_0$. It can be justified by averaging theory \cite{Sanders-Verhulst-Murdock-Spinger_2007} thanks to inequalities \eref{eq:orders},\eref{eq:contineq}.
Within this approximation, $\ket{\phi}$ follows the dynamics
\begin{equation}
\label{eq:dyn_phi}
i \dotex \ket{\phi(t)}=(\bar{H}_I+\widetilde{H}_I(t))\, \ket{\phi(t)}
\end{equation}
where
\begin{eqnarray*}
\bar{H}_I&=&\sum_{k=0}^{N-1}{k\Delta_k \, \proj{k}}\\
\widetilde{H}_I(t)&=&\sum_{k=0}^{N-2}{\mu_k(E(t)\ket{k}\bra{k+1}+E^*(t)\ket{k+1}\bra{k})}\; .
\end{eqnarray*}


\subsection{Change of frame}\label{ss:cv}

Hamiltonian $\widetilde{H}_I(t)$ contains a control field whose phase $\frac{1}{\varepsilon}\theta(\varepsilon t)$ varies on timescales of order one.
The key idea to apply adiabatic passage to the $N$-level system is an appropriate further change of frame on \eref{eq:dyn_phi}, such that all explicit time-dependence in the resulting dynamics involves timescales of order $\varepsilon$.
To this end, we extend the change of frame given in \cite[Section 4.6]{Allen-Eberly_1987} for the two-level case and define $\; \ket{\xi(t)} = \; \sum_{k=0}^{N-1} \; e^{k \frac{i}{\varepsilon}\theta(\varepsilon t)} \, \proj{k} \; \ket{\phi(t)}$.
Dynamics \eref{eq:dyn_phi} becomes
\begin{equation}
\label{eq:dyn_xi}
i \dotex \ket{\xi(t)}=(H_R(\omega(\varepsilon t))+A(\varepsilon t)H_1)\, \ket{\xi(t)}
\end{equation}
with $\omega=\theta'$, $H_1$ given by \eref{eq:H1} and
\begin{equation}\label{eq:Lambda}
H_R(v) = \sum_{k=0}^{N-1}{k(\Delta_k-v) \, \proj{k}} \;\; \mbox{ for all } v\in\RR \; .
\end{equation}
Define the propagator $U^\varepsilon$ to be a time-dependent $N$ by $N$ unitary matrix such that the solution of \eref{eq:dyn_xi} is given by
$\ket{\xi(t)}=U^\varepsilon(t)\ket{\xi(0)}$ for all $t$ and for all $\ket{\xi(0)}$. Then $U^\varepsilon$ follows the dynamics
\begin{eqnarray}
\label{eq:dyn_U}
i\varepsilon \frac{d}{ds} U^\varepsilon(s) & = & H(s)\, U^\varepsilon(s)\;\; ,\quad U^\varepsilon(0) = I\\[1mm]
\label{eq:hamiltonians}
\mbox{with} & & H(s) \, = \, H_R(\omega(s))+A(s)H_1
\end{eqnarray}
in the time scale $s=\varepsilon t$.
In the following, we study system \eref{eq:dyn_U} for $s$ in the interval $[0,1]$ and with $A(s)$ and $\omega(s)$ as controls.
Our goal is to achieve: \begin{enumerate*}
\item[(a)] \emph{Adiabatic approximate eigenstate permutations:}
\begin{equation}\label{eq:goal}
\!\!\! \lim_{\varepsilon\rightarrow 0^+}\; \max_{k \in G} \,\Vert\, U^\varepsilon(1) \proj{k} U^\varepsilon(1)^\dag - \proj{\sigma(k)}\, \Vert = 0
\end{equation}
for given $G \subseteq \NN_0^{N\minou 1}$ and given permutation $\sigma$ of $(0,\ldots ,N-1)$.
\item[(b)] \emph{Ensemble control:} a single control $(A,\omega)$ achieves such eigenstate permutation on an ensemble of systems with different parameter values; the parameters are the dipole moments $(\mu_0,\ldots,\mu_{N-2})$ and, in some cases, the anharmonicities $(\Delta_0,\ldots,\Delta_{N-1})$.
\item[(c)] \emph{Robust control inputs:} the above holds for any $(A,\omega)$ that satisfy a set of key properties.
\end{enumerate*}
\begin{rmk}
Writing \eref{eq:goal} in terms of $\proj{k}$, the projector on eigenspace $\{ \beta \, \ket{k} : \beta \in \CC \}$, expresses that the goal is really population transfer, i.e.~we allow $U^\varepsilon(1) \ket{k} \approx e^{i \chi_k} \ket{\sigma(k)}$ with arbitrary phases $\chi_k \in \SScirc$. Both frame changes --- for RWA in section \ref{ss:standard} and $\theta$-dependent in section \ref{ss:cv} --- involve only phase changes on eigenstates. Therefore, for all $t$ and for all $\ket{k}$,
\begin{eqnarray*}
	\Vert \proj{\psi(t)} - \proj{k} \Vert & = & \Vert \proj{\phi(t)} - \proj{k} \Vert\\
	& = & \Vert \proj{\xi(t)} - \proj{k} \Vert \; .
\end{eqnarray*}
\end{rmk}


\section{Robust ensemble transfer from $\ket{k}$ to $\ket{N-k-1}$}\label{sec:0toN}

In this section we consider a control protocol -- chirped pulse -- used by physicists to drive a system from the lowest eigenspace, spanned by $\ket{0}$, to the highest eigenspace, spanned by $\ket{N-1}$, of the free Hamiltonian $H_0$ given in \eref{eq:H0}. In fact we prove that a general (robust) class of control inputs transfers population from eigenstate $\ket{k}$ to eigenstate $\ket{N-k-1}$, for all $k$, on an ensemble of systems with different values of parameters $\mu_0,\dots,\mu_{N-2}$ (dipole moments) and $\Delta_0,\ldots,\Delta_{N-1}$ (anharmonicities).

The key requirements on the control are (i) to use a sufficiently chirped pulse --- condition (b) in Theorem \ref{thm:1} --- and (ii) to avoid all eigenvalue crossings --- condition (c) in Theorem \ref{thm:1}.

\subsection{Transfer Theorem}

For $k=0,\ldots,N-1$ let $\lambda^R_k(s) = \bra{k}H_R(\omega(s))\ket{k}=k(\Delta_k-\omega(s))$, the eigenvalues of $H_R(\omega(s))$.

\begin{thm}\label{thm:1}
For given $\Delta>0$, $\mu_{max}>\mu_{min}>0$, consider $\SSS$ an ensemble of systems of type \eref{eq:dyn_U} with $\mu_j \in [\mu_{min},\mu_{max}]$ for all $j \in \NN_0^{N\minou 2}$ and $\Delta_j \in [-\Delta,\Delta]$ for all $j \in \NN_0^{N\minou 1}$.
Take controls $A$ and $\omega$ with:
\begin{enumerate}
\item[(a)] $A$ and $\omega$ $\in \mathcal{C}^2([0,1],\RR)$
\item[(b)] $\omega(0)$ and $\omega(1)$ are such that, for all systems in $\SSS$,
\begin{eqnarray}\label{eq:ineqeigen}
& & \lambda^R_0(0)< \ldots <\lambda^R_{N-1}(0)\;\;\; \mbox{\small and }\\
\nonumber & & \lambda^R_0(1)> \ldots >\lambda^R_{N-1}(1)
\end{eqnarray}
\item[(c)] $A(0)=A(1)=0$ and $A(s)\neq 0$ for $s\in]0,1[$
\end{enumerate}
Then $\exists$ a constant $C>0$ such that, for all $\varepsilon >0$,
$$\mathop{\sup_{\SSS}}_{k \in \NN_0^{N\minou 1}}\!\!\! \Vert \; U^\varepsilon(1)\proj{k}U^\varepsilon(1)^\dag-\proj{N-k-1} \; \Vert \;\;\; \leq \;\; C \varepsilon \, .$$
\end{thm}
The proof of this theorem is given in section~\ref{sec:proofs}; we there actually replace the simple condition (c) on $A$ by a more general one: $A(0)=A(1)=0$
and $A(s)\ne 0$ for all $s\in\mathcal{I}^\omega(\SSS)$, where
\begin{eqnarray}
\label{eq:singularpointset} \mathcal{I}^\omega & = & \{s\in[0,1]: H_R(\omega(s)) \mbox{ \emph{is degenerate for some system} } \in \SSS \}\, .
\end{eqnarray}
The argument is based on the facts that the system approximately follows eigenstates of $H(s)$ for small enough $\varepsilon$ (adiabatic theorem), eigenvalues of $H_R$ are inverted between $s=0$ and $s=1$ thanks to $\omega(s)$ (chirping), and nonzero $A(s)$ avoids all crossings for eigenvalues of $H(s)$ such that e.g.~the initial highest-energy level $\ket{N-1}$ connects to the final highest-energy level $\ket{0}$ (see Lemma~\ref{Lemma:nondegeneracy} in section~\ref{sec:proofs}).
Theorem \ref{thm:1} implies that for a given control satisfying the assumptions, taking $\varepsilon$ small enough allows to invert the state populations of a whole ensemble of systems featuring different parameter values. The control inputs only need to satisfy a few weak conditions and are therefore robust to many perturbations.
These insensitivity properties of the adiabatic passage protocol have long been recognized by experimentalists. They commonly use the following type of control, see e.g.~\cite{Vitanov01}.

\begin{example}\label{ex:chirp}
A function $\omega$ satisfying the inequalities \eref{eq:ineqeigen} is e.g.~$\omega(s)=\alpha(s-\frac{1}{2})$, for a large enough positive $\alpha$; such $\omega$ is said to perform a frequency sweep.
Except for the finite extension of time domain, such inputs are obtained by a ``chirped'' Gaussian laser pulse, which takes the form $E(t) = E_0\, \int_{-\infty}^{+\infty} e^{-\zeta^2 \tau^2}\, e^{i \kappa \zeta^2}\, e^{-i \zeta t}  \, d\zeta$ where $\kappa\neq 0$ characterizes chirping.
\end{example}
Theorem \ref{thm:1} still holds if inequality \eref{eq:ineqeigen} is replaced by
\begin{equation*}
\lambda^R_0(0) > \ldots > \lambda^R_{N-1}(0)\; \mbox{ and } \lambda^R_0(1)< \ldots <\lambda^R_{N-1}(1)\; ,
\end{equation*}
i.e.~the direction of the frequency sweep in Example \ref{ex:chirp} can be inverted (taking a large enough \emph{negative} $\alpha$). However, for a given system, choosing one inequality over the other may allow to get a lower value for the constant hidden in the ``order of magnitudes'' result.
This brings a mathematical foundation to the experimental observations made e.g.~in \cite{Maas-al-Noordam-PRA_1999}.


\subsection{Simulations}\label{subsection:simu1}

\begin{figure}[th]
\begin{center}	
\includegraphics[width=5in]{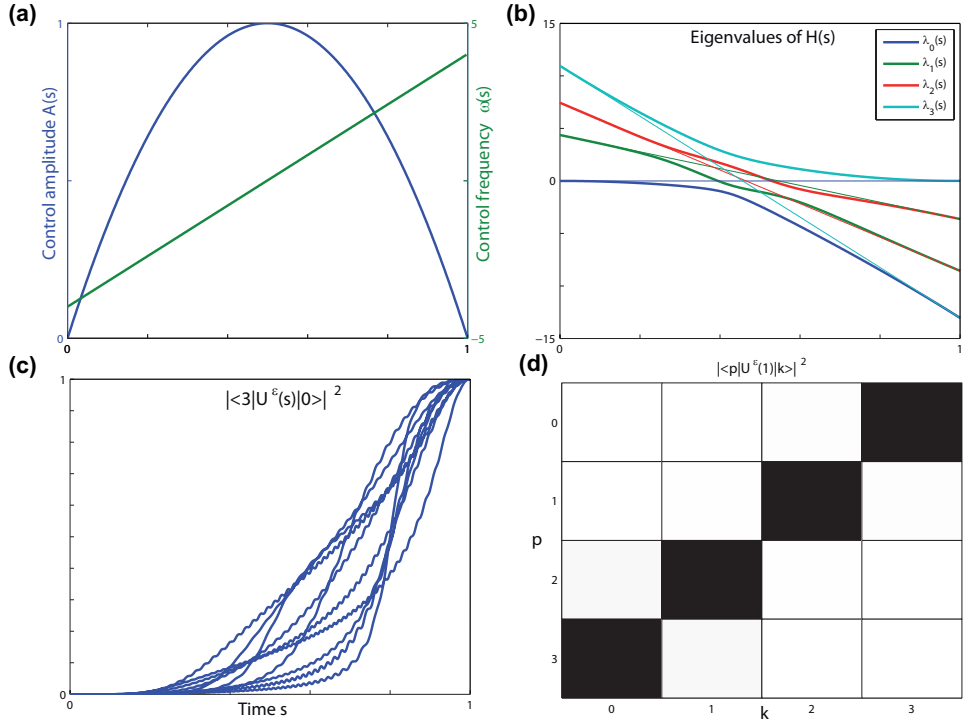}
\caption{Control scheme transferring $\ket{k}$ to $\ket{N-k-1}$. (a) control inputs $A(s)$, $\omega(s)$. (b) $s$-dependent eigenvalues of $H(s)$ (thick lines) and of $H_R(\omega(s))$ (thin lines). (c) population on level $\ket{3}$ for 10 systems whose parameters $\mu_0,\mu_1,\mu_2$ and $\Delta_1,\Delta_2,\Delta_3$ were randomly picked respectively in $[1,5]$ and $[-0.4,0.4]$, and all starting at the initial state $\ket{0}$. (d) squared norm of the matrix elements of $U^\varepsilon(1)$, represented in shading from white (value 0) to black (value 1).}
\label{fig:fig1}
\end{center}
\end{figure}

We simulate system \eref{eq:dyn_U} with a control satisfying assumptions (a), (b) and (c) of Theorem \ref{thm:1}.
We consider a $4$-level quantum ladder (so $N=4$). We take $\varepsilon=10^{-2}$,
$\Delta_0,\ldots,\Delta_3 \in [-0.4,0.4]$ and $\mu_0,\mu_1,\mu_2 \in [\mu_{min},\mu_{max}]=[1,5]$.
The control is $\omega(s)=8(s-\frac{1}{2})$ and $A(s)=s(1-s)$,
represented on Fig.\ref{fig:fig1}.a.
Fig.\ref{fig:fig1}.b shows how the eigenvalues of $H(s)$ (thick lines) avoid crossing
For the illustrated random choice of detunings, the eigenvalues of $H_R(\omega(s))$ (thin lines) are very close to concurrent between $s=0.5$ and $s=0.6$. This poses no problem for the adiabatic transfer from $\ket{k}$ to $\ket{N-k-1}$. The successful transfer is illustrated on Fig.\ref{fig:fig1}.d, which shows the squared norm of the projection of
$U^{\varepsilon}(1) \ket{k}$ onto $\ket{p}$, for all pairs $(\ket{k},\ket{p})$ of eigenvectors of $H_0$; this is equivalent to the squared norm of element on
row $p$, column $k$ of matrix $U^{\varepsilon}(1)$ that acts by left-multiplication on initial column-vectors, for $U^{\varepsilon}(1)$ expressed in basis $(\ket{0},\ldots,\ket{3})$. Fig.\ref{fig:fig1}.c shows ensemble control on ten systems with different random values of
$\Delta_0,\ldots,\Delta_3$ and $\mu_0,\mu_1,\mu_2$.


\section{Robust ensemble transfer from $\ket{l}$ to $\ket{p}$}\label{sec:ltop}

In this section we propose a new robust control protocol to drive a system from the eigenspace (of free Hamiltonian $H_0$) spanned by $\ket{l}$ to the eigenspace spanned by $\ket{p}$, for any given $l$ and $p$ in $\NN_0^{N\minou 1}$. The population transfer works on an ensemble of systems with different values of $\mu_0,\dots,\mu_{N-2}$ (dipole moments), and for a general class of inputs where zero-crossings of $A(s)$ must be correlated with degeneracies of $H_R(\omega(s))$; the latter depend on $\omega(s)$ and (some of the) anharmonicities $\Delta_0,\ldots,\Delta_{N-1}$, which must hence be fixed.


\subsection{From $\ket{0}$ to any $\ket{p}$}\label{ss:0top}

For the sake of clarity, we start by giving sufficient conditions on $A$ and $\omega$ for the particular population transfer from $\ket{0}$ to arbitrary level $\ket{p}$.
Section \ref{ss:ltop} generalizes the result to arbitrary initial state $\ket{l}$. Consider the following assumptions:
\begin{enumerate}
\item[(A1)] $\SSS$ is an ensemble of systems of type \eref{eq:dyn_U} with $\mu_j \in [\mu_{min},\mu_{max}]$ for all $j \in \NN_0^{N\minou 2}$, for some given $\mu_{max} > \mu_{min} > 0$, and with given sequence of detunings $(\Delta_0,\ldots,\Delta_{N-1})$, such that the set $\{k(\Delta_k-v)\colon k\in \NN_0^{N\minou 1}\}$ contains at least $N-1$ distinct values for any $v\in\RR$;
\item[(A2)] $\omega$ is analytic and $\frac{d}{ds}\omega(s)>\gamma>0$ for all $s\in[0,1]$;
\item[(A3)] $\omega(0)$ and $\omega(1)$ are such that \eref{eq:ineqeigen} holds.
\end{enumerate}
For any $m$ and $n$ in $\NN_0^{N\minou 1}$ with $m<n$, we note $s(m,n)$ the unique time\footnote{If assumptions (A1) to (A3) hold, then the existence and unicity of $s(m,n)$ is ensured for all $m$ and $n>m$:
see Fig.\ref{fig:fig2}.b or Fig.\ref{fig:fig3}.c}
where $\lambda^R_m(s(m,n)) = \lambda^R_n(s(m,n))$.

As all systems in $\SSS$ have the same sequence of detunings, they feature the same eigenvalues $\lambda^R_0,\ldots,\lambda^R_{N-1}$ of $H_R$ and hence the same set of $s(m,n)$.
The set of all $s(m,n)$ equals $\mathcal{I}^\omega$ defined in \eref{eq:singularpointset}, with dependence on particular system $\in \SSS$ becoming irrelevant.
The end of (A1) further implies that $H_R$ has at most one pair of equal eigenvalues for any $s \in [0,1]$ i.e.~$(m,n) \neq (j,k)$ implies $s(m,n) \neq s(j,k)$, hence $\mathcal{I}^\omega$ contains $N(N-1)/2$ distinct values.
Further define $\mathcal{I}^\omega_0 = \{ s_1,\ldots,s_{N-1} \} \subset \mathcal{I}^\omega$ the $N-1$ points where $\lambda_0^R(s) = \lambda_n^R(s)$ for some $n \in \NN_1^{N\minou 1} $, numbered such that $s_1<s_2<\ldots<s_{N-1}$. Thus, for each $s_k\in\mathcal{I}^\omega_0$ there exists a unique $n\in\NN_1^{N\minou 1}$ such that $s_k=s(0,n)$.

The key requirements on the control to achieve population transfer from $\ket{0}$ to $\ket{p}$ are (i) to use a sufficiently chirped pulse frequency --- condition (A3) --- and (ii)
to shape pulse amplitude in order to appropriately provoque --- (c) in Theorem \ref{thm:2} --- or avoid --- (b),(d) in Theorem \ref{thm:2} --- crossing of eigenvalues of $H$.

\begin{thm}
\label{thm:2}
Consider $\SSS$ an ensemble of systems satisfying (A1) with a control $\omega$ satisfying (A2) and (A3). Take
$p\in\{0,\ldots,N-1\}$
and consider a control $A$ with the following properties:
\begin{enumerate}
\item[(a)] $A$ is analytic over $[0,1]$ and $A(0)=A(1)=0$.
\item[(b)] $A(s)\neq 0$ for all $s\in \mathcal{I}^\omega\backslash \mathcal{I}^\omega_0$.
\item[(c)] $A(s_k)= 0$ for all $s_k \in \mathcal{I}^\omega_0$ with $k \leq N-p-1$.
\item[(d)] $A(s_k)\neq 0$ for all $s_k \in \mathcal{I}^\omega_0$ with $k \ge N-p$.
\end{enumerate}
Then $\exists$ a constant $C>0$ such that, for all $\varepsilon >0$,
$$\;\;\; \sup_{\SSS} \; \Vert \; U^\varepsilon(1)\proj{0}U^\varepsilon(1)^\dag-\proj{p} \; \Vert \;\;\; \leq \;\; C \sqrt{\varepsilon} \, .$$
\end{thm}
The proof, given in section~\ref{sec:proofs}, shows that at eigenvalue crossing points the system adiabatically follows the eigenvector corresponding to the \emph{crossing} branch.


\subsection{From any $\ket{l}$ to any $\ket{p}$}\label{ss:ltop}

Under assumptions (A1) to (A3), we denote $\mathcal{I}^\omega_{k+}(s)=\{ s(m,n) \in \mathcal{I}^\omega : m=k,\; n>k \mbox{ and } s(m,n) > s \}$ and $\; \mathcal{I}^\omega_{k-}(s)=\{ s(m,n) \in \mathcal{I}^\omega : m<k,\; n=k \mbox{ and } s(m,n) > s \}$, for any $k \in \NN_0^{N\minou 1}$. Further let $q_{k\pm}(s)=\inf(\mathcal{I}^\omega_{k\pm}(s))$ and define $g_{k\pm}(s)$ by
$s(k,g_{k+}(s)) = q_{k+}(s)$ and $s(g_{k-}(s),k) = q_{k-}(s)$ respectively.
For $p \leq N-l-1$, construct $\mathcal{I}^\omega_{lp}$ with the following algorithm.
\begin{algorithmic}[1]
\STATE $d:=0$; $x:=0$; $k:=l$; $\mathcal{I}^\omega_{lp}:=\emptyset$;
\WHILE{$d < N-l-p-1$}
\WHILE{[$\; \mathcal{I}^\omega_{k-}(x)\neq \emptyset$ \mbox{\textbf{ and }} $q_{k-}(x) < q_{k+}(x)\;$]}
\STATE $k:= g_{k-}(x)$; $x:=q_{k-}(x)$;
\ENDWHILE
\STATE $\mathcal{I}^\omega_{lp}:=\mathcal{I}^\omega_{lp} \cup \{q_{k+}(x) \}$; $d:=d+1$; $x:=q_{k+}(x)$;
\ENDWHILE	
\end{algorithmic}
The algorithm is verified to always successfully complete\footnote{Indeed by construction, the cardinality of $\mathcal{I}^\omega_{k+}(x)$ equals $N-l-d-1$ (except \emph{during} the update on line 6) and the cardinality of $\mathcal{I}^\omega_{k-}(x)$ decreases by one each time line 4 is applied; thus it is impossible to keep applying line 4 infinitely, and line 6 is always well-defined (that is $\mathcal{I}^\omega_{k+}(x) \neq \emptyset$) for $d < N-l-1$.}.
For $p \geq N-l-1$, we can define $\mathcal{I}^\omega_{lp}$ with a similar algorithm but where `$<$' is changed to `$>$' on line 2 and indices $\vphantom{k}_{k-}$, $\vphantom{k}_{k+}$ are switched. Then $\mathcal{I}^\omega_{lp}$ contains $\vert N-l-p-1 \vert$ elements.

\begin{crl}
\label{crl:1}
Consider $\SSS$ an ensemble of systems satisfying (A1) with a control $\omega$ satisfying (A2) and (A3). Take
$l,p$ in $\NN_0^{N\minou 1}$ and consider a control $A$ with the following properties:
\begin{enumerate}
\item[(a)] $A$ is analytic over $[0,1]$ and $A(0)=A(1)=0$.
\item[(b)] $A(s)=0$ for all $s\in \mathcal{I}^\omega_{lp}$.
\item[(c)] $A(s)\neq0$ for all $s\in \mathcal{I}^\omega\setminus \mathcal{I}^\omega_{lp}$.
\end{enumerate}
Then $\exists$ a constant $C>0$ such that, for all $\varepsilon >0$,
$$\;\;\; \sup_{\SSS} \; \Vert \; U^\varepsilon(1)\proj{l}U^\varepsilon(1)^\dag-\proj{p} \; \Vert \;\;\; \leq \;\; C \sqrt{\varepsilon} \, .$$
\end{crl}

Assumption (A1) ensures that each eigenvalue crossing / anti-crossing can be addressed individually. This ensures that any transfer can be implemented in any situation, but it is in general not necessary for a given system and transfer, as (simultaneous) crossings of some eigenvalue branches are irrelevant. The control proposed for Theorem \ref{thm:2} or Corollary \ref{crl:1} is just one amongst many possibilities of ``eigenvalue crossing designs''. Indeed, depending on $(l,p)$ and on the particular arrangement of the $s(m,n)$, one can find other subsets $\mathcal{J}_{lp} \subset \mathcal{I}^\omega$ such that taking $A(s)=0$ if and only if $s \in \mathcal{J}_{lp}$, permutes the eigenvalues in such a way that $\lambda_l(1)=\lambda^R_p(1)$. The controls that we propose are optimal in the sense that they require a minimal number of pairwise crossings, that is of annihilations of $A$ at accurate points. Variant annihilation subsets $\mathcal{J}_{lp}$ may be useful (i) to avoid some crossing points $s(m,n)$ or eigenvalue branches (e.g.~because corresponding $\Delta_m$ or $\Delta_n$ is poorly known, or because $s(m,n)$ is close to some other point in $\mathcal{I}^\omega$), (ii) to optimize adiabatic convergence as a function of $\varepsilon$,
or (iii) to simultaneously perform population transfers between several eigenstates, as we do in section \ref{sec:permutations}.\\

Another approach \cite{Thomas-Guerin-Jauslin-PRA_2005} for transferring $\ket{l}$ to $\ket{p}$ is to use $A(s)$ Gaussian, i.e.~without any annihilations, but reduce $\omega(s)$ to a specific range. Indeed, under the above assumptions, it is possible to choose $\omega_{\mbox{min}}$ and $\omega_{\mbox{max}}$ such that $l (\Delta_l-\bar{v}) = p(\Delta_p-\bar{v})$ for some $\bar{v} \in [\omega_{\mbox{min}},\omega_{\mbox{max}}]$ and $H_R(v)$ is non-degenerate for all $v \in [\omega_{\mbox{min}},\omega_{\mbox{max}}] \setminus \{\bar{v}\}$. Then taking $\omega(s)$ monotone between $\omega_{\mbox{min}}$ and $\omega_{\mbox{max}}$ just induces one avoided crossing that exchanges $\ket{l}$ and $\ket{p}$. Pictorially, this is like selecting a particular narrow vertical slice on Fig.\ref{fig:fig1}.b. Depending on the specific system under study and whether it is experimentally easier to precisely modulate the amplitude or the phase of a field, one method may be more suitable than the other. A main advantage of our method is that, unlike the method proposed in \cite{Thomas-Guerin-Jauslin-PRA_2005}, it can be extended to achieve any permutation of eigenstates as is shown in section \ref{sec:permutations}.


\subsection{Simulations}

\begin{figure}[th]
\begin{center}	
\includegraphics[width=5in]{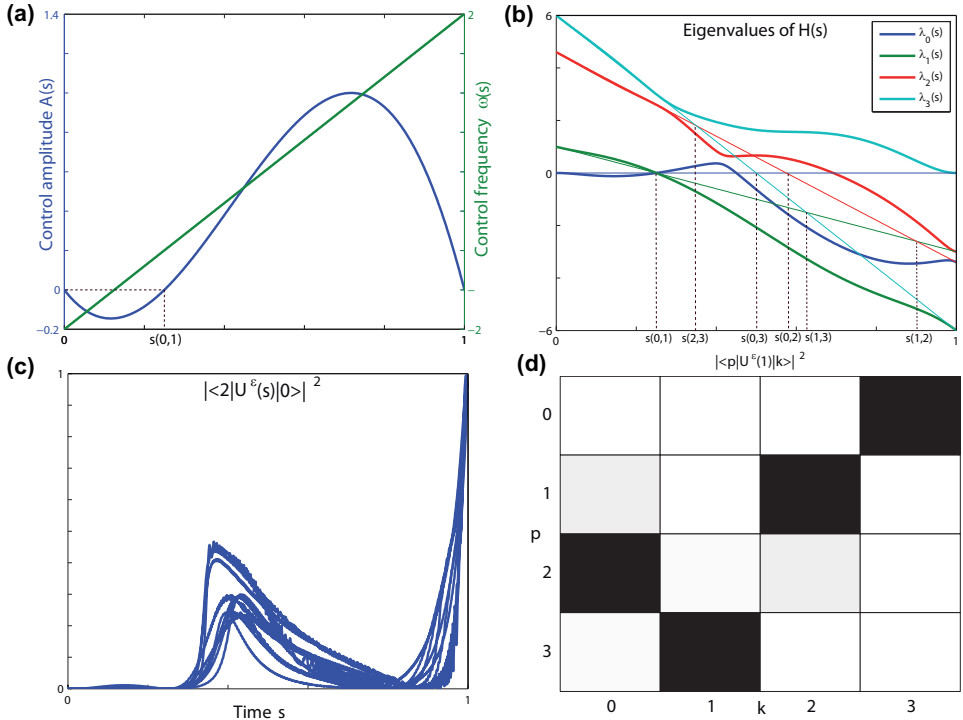}
\caption{Control scheme transferring $\ket{0}$ to $\ket{2}$ control scheme; subplots analogue to Fig.\ref{fig:fig1}, except that $\Delta_1,\Delta_2,\Delta_3$ remain fixed for (c). A(s) vanishes at $s=0.25$ so that $\lambda_0$ and $\lambda_1$ cross instead of avoiding crossing.}
\label{fig:fig2}
\end{center}
\end{figure}

As in section \ref{subsection:simu1} we simulate \eref{eq:dyn_U} for a $4$-level quantum ladder (so $N=4$) with $\mu_0,\mu_1,\mu_2 \in [\mu_{min},\mu_{max}]=[1,5]$.
We now take $\varepsilon=10^{-3}$ and in accordance with the statement of Theorem \ref{thm:2} we fix the anharmonicities, taking $\Delta_1=-1$, $\Delta_2=0.3$, $\Delta_3 = 0$ (the value of $\Delta_0$, multiplied by $k=0$, is irrelevant).
We target in particular a transfer from $\ket{0}$ to $\ket{2}$. The algorithm of section \ref{ss:ltop} reduces to the simple case of Theorem \ref{thm:2}, requesting a single zero of $A(s)$ at $s=\inf\{s(0,1),s(0,2),s(0,3)\} = s(0,1) = 0.25$ in addition to $A(0)=A(1)=0$. We take $A(s)=s(1-s)(s-0.25)$ and $\omega(s)=4(s-\frac{1}{2})$, represented on Fig.\ref{fig:fig2}.a. Fig.\ref{fig:fig2}.b shows how the eigenvalues $\lambda_k(s)$ of $H(s)$ cross or not (thick lines);
the eigenvalues of $H_R(\omega(s))$ (thin lines) define points $s(m,n)$ for our control design.
Fig.\ref{fig:fig2}.d confirms achievement of the intended result by showing the squared norm of components of matrix $U^{\varepsilon}(1)$ in basis $(\ket{0},\ldots,\ket{3})$: we indeed have $\vert \langle p \vert U^{\varepsilon}(1) \vert k \rangle \vert^2 \approx 1$ for $(p,k)=(2,0)$ (other values incidental). Fig.\ref{fig:fig2}.c illustrates ensemble control on ten systems with different random values of $\mu_0,\mu_1,\mu_2$.
Since for this particular case the control only exploits precise crossing point $s(0,1)=0.25$, we might actually allow ensembles with different $\Delta_2,\Delta_3$.


\section{Robust ensemble permutation of populations}\label{sec:permutations}

In this section we describe the most general result of the paper, adiabatically transferring $(\proj{0},\ldots,\proj{N-1})$ to $(\proj{\sigma(0)},\ldots,\proj{\sigma(N-1)})$, where $\sigma$ is any permutation of $\NN_0^{N\minou 1}$. As in section \ref{sec:ltop}, the population permutation works on an ensemble of systems with different values of $\mu_0,\dots,\mu_{N-2}$ (dipole moments), and for a general class of inputs where zero-crossings of $A(s)$ must be correlated with degeneracies of $H_R(\omega(s))$; the latter depend on $\omega(s)$ and require anharmonicities $\Delta_0,\ldots,\Delta_{N-1}$ to be fixed and known.
We prove existence of an appropriate control by recurrence on $N$. In fact this recurrence method can be used to design $A(s)$, as we illustrate in section \ref{subsec:simu3}.


\subsection{Permutation theorem}

\begin{thm}
\label{thm:3}
Consider $\SSS$ an ensemble of systems satisfying (A1) with a control $\omega$ satisfying (A2) and (A3). Take $\sigma$ any permutation of $\NN_0^{N\minou 1}$.
Then there exists a subset $\mathcal{I}_A \subseteq \mathcal{I}^\omega$ for which, taking control $A$ to satisfy
\begin{enumerate}
\item[(a)] $A$ analytic over $[0,1]$ and $A(0)=A(1)=0$,
\item[(b)] $A(s)=0$ for all $s\in \mathcal{I}_A$,
\item[(c)] $A(s)\neq 0$ for all $s\in \mathcal{I}^\omega\setminus \mathcal{I}_A$,
\end{enumerate}
implies: $\exists$ a constant $C>0$ such that, for all $\varepsilon >0$,
$$\mathop{\sup_{\SSS}}_{k \in \NN_0^{N\minou 1}}\! \Vert \; U^\varepsilon(1)\proj{k}U^\varepsilon(1)^\dag-\proj{\sigma(k)} \; \Vert \;\;\; \leq \;\; C \sqrt{\varepsilon} \, .$$
\end{thm}

Since the proof of this Theorem is constructive and necessary for the understanding of the example below, we present it here.

\begin{pf}[of Theorem \ref{thm:3}]
The formal arguments (sup, adiabatic propagator) are presented in detail in the proof of Theorem \ref{thm:1} in section \ref{sec:proofs}. We focus on the construction of the control $A(s)$ by following analytic eigenvalue branches of $H(s)$.
	
The property is obvious for $N=2$: either $(\sigma(0),\sigma(1)) = (1,0)$, which follows Theorem \ref{thm:1} just requiring $A(s(0,1)) \neq 0$; or $(\sigma(0),\sigma(1)) = (0,1)$, which follows Theorem \ref{thm:2} transferring $\ket{0}$ to $\ket{p}=\ket{0}$ with one crossing\footnote{Indeed, $\{ P_{\lambda_0(1)},P_{\lambda_1(1)} \}=\{\proj{0},\proj{1}\}$ then automatically implies transferring $\proj{1}$ to $P_{\lambda_1(1)}=\proj{1}$.}, i.e.~just requiring $A(s(0,1))=0$.

Assume that we can achieve any permutation of $\NN_0^{K\minou 1}$ for $N=K$, and we are given a permutation $\sigma$ of $\NN_0^{K}$ for $N=K+1$ where $\sigma(l)=K$ and $\sigma(K)=p$.
\newline $\bullet$ If $l=p=K$, i.e. $\sigma(K)=K$, then first build the remaining permutation on levels $\ket{0},\ldots,\ket{K-1}$ by neglecting level $\ket{K}$. This uses the result for $N=K$; it just requires $A(s)=0$ for some $s = s(m,n)$ and $A(s)\neq 0$ for some other $s =s(m,n)$, with $m,n < K$. Now take a particular such $A(s)$ where in addition, $A(s)=0$ for all $s \in \{s(m,K): m \in \NN_0^{K\minou 1}\}$. Then $\lambda_K(s)$, starting at $\lambda_K(0) = \lambda_K^R(0)$, exactly follows the same crossings as $\lambda_K^R(s)$ to end up as $\lambda_K(1)=\lambda_K^R(1)$; the other levels remain unperturbed, so $\sigma$ is achieved.
\newline $\bullet$ If $l \neq K \neq p$, then first construct $\overline{A}(s)$ by applying the result of the preceding point to $\overline{\sigma}$, defined by
$$\overline{\sigma}(l)=p \, ; \;\; \overline{\sigma}(K)=K \, ; \;\; \overline{\sigma}(k)=\sigma(k) \mbox{ for all } k\not\in \{ l,K\} \, .$$
$\overline{A}(s)$ performs the target permutation, except that $K$ remains on $K$ and $l$ goes to $p$. From (\ref{eq:ineqeigen}) eigenvalue branch $\lambda_K(s)$ necessarily crosses, at some $\overline{s} \in \{s(m,K): m \in \NN_0^{K\minou 1}\}$, the analytic eigenvalue branch that starts at $\lambda_l(0)=\lambda^R_l(0)$ and ends at $\lambda_l(1)=\lambda^R_{p}(0)$. Define $A(s)$ to have the same zeros as $\overline{A}(s)$ except that $A(\overline{s}) \neq 0$. This just transforms the crossing at $\overline{s}$ into an anti-crossing, such that the analytic branch coming from $\lambda_K(0)$ (resp.~$\lambda_l(0)$) now connects to the analytic branch going to $\lambda_{p}(1)$ (resp.~$\lambda_K(1)$). Thus $A(s)$ achieves the target permutation $\sigma$.\hfill $\boldsymbol{\Box}$
\end{pf}

Each ``eigenvalue crossing design'' choice $\mathcal{I}_A$ yields a particular permutation $\sigma_{\mathcal{I}_A}$. For $N>2$, the number $2^{N(N-1)/2}$ of possible $\mathcal{\mathcal{I}_A}$ (i.e.~subsets of $\mathcal{I}^\omega$) is strictly larger than the number $N!$ of permutations. Thus there are still several $\mathcal{\mathcal{I}_A}$ that yield the same $\sigma$. Unlike in section \ref{sec:ltop}, building $A(s)$ as in the proof of Theorem \ref{thm:3} does not necessarily yield a minimal cardinality of $\mathcal{\mathcal{I}_A}$ for given $\sigma$.



\subsection{Example and simulations}

\label{subsec:simu3}
\begin{figure}[th]
\begin{center}	
\includegraphics[width=5in]{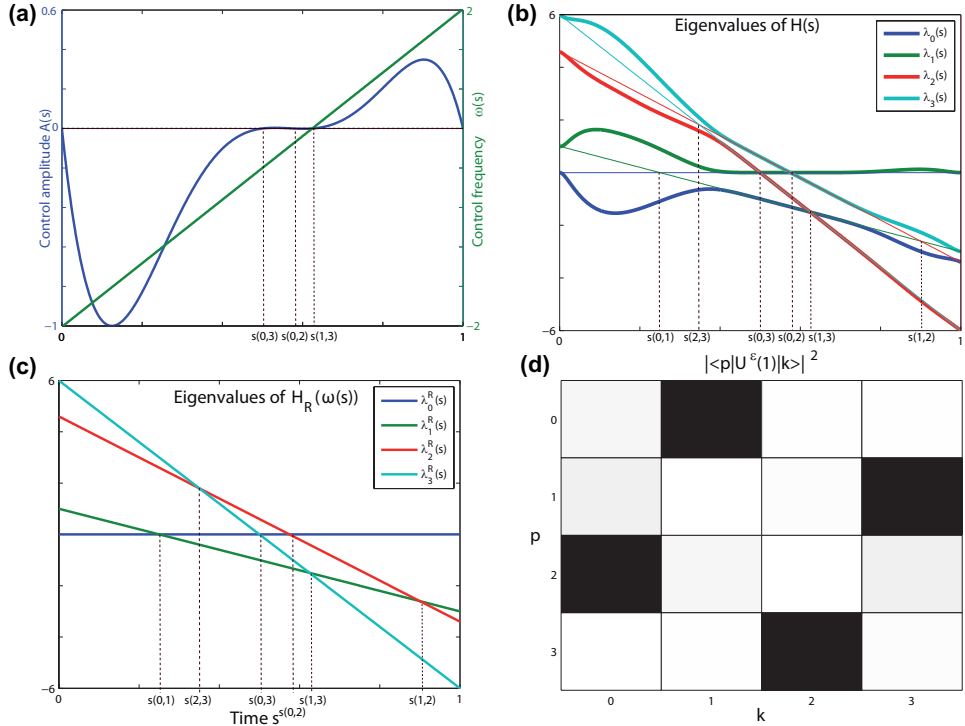}
\caption{Control scheme generating permutation $\sigma(0,1,2,3)=(2,0,3,1)$ and simulation result. Subplots (a),(b),(d) analogue to Fig.\ref{fig:fig1}). Subplot (c) shows the eigenvalues of $H_R(\omega(s))$, used to design $\mathcal{I}_A$ (see text). The points $s \in \mathcal{I}_A$ where $A(s)=0$ are marked on (a).}
\label{fig:fig3}
\end{center}
\end{figure}

We first illustrate the control design by recurrence based on the proof of Theorem \ref{thm:3}. Consider target permutation $\sigma(0,1,2,3)=(2,0,3,1)$. First we reduce it down to an elementary permutation. Start with $K=N-1=3$ and note $(l,p) = (2,1)$ because $\sigma(2)=K$ and $\sigma(K)=1$; we thus define $\overline{\sigma}(0,1,l=2,3)=(2,0,p=1,3)$ and impose $\overline{A}(s)=0$ for $s \in \{s(0,3),s(1,3),s(2,3)\}$ reducing the permutation to $0,1,2$.
Then we take $\overline{K}=N-1=2$ and note $(\overline{l},\overline{p})=(0,1)$ because $\overline{\sigma}(0)=\overline{K}$ and $\overline{\sigma}(\overline{K})=1$; we thus define $\overline{\overline{\sigma}}(\overline{l}=0,1,2,3)=\overline{\overline{\sigma}}(\overline{p}=1,0,2,3)$ and impose $\overline{\overline{A}}(s)=0$ for $s \in \{s(0,2),s(1,2) \}$ reducing the permutation to $0,1$. To implement $\overline{\overline{\sigma}}$ we need $\overline{\overline{A}}(s(0,1)) \neq 0$. Now we progressively move up to permutations on more levels, removing one $\overline{\phantom{a}}$ at a time from our objects. The reader is encouraged to follow crossings/anti-crossings under the different controls by referring to Fig.\ref{fig:fig3}.c,
corresponding to our chirping choice $\omega(s)=4(s-\frac{1}{2})$.
Under $\overline{\overline{A}}$ the analytic branch from $\ket{\overline{l}} = \ket{0}$ to $\ket{\overline{p}}=\ket{1}$ meets the branch staying on $\ket{\overline{K}}=\ket{2}$ at $\overline{\overline{s}}=s(1,2)$. We therefore impose $\overline{A}(s(1,2))\neq 0$ unlike for $\overline{\overline{A}}$, and for the rest copy the requirements of $\overline{\overline{A}}$: $\overline{A}(s(0,1))\neq 0$, $\overline{A}(s(0,2)) = 0$. Now under $\overline{A}$ the branch from $\ket{l}=\ket{2}$ to $\ket{p}=\ket{1}$ crosses the branch staying on $\ket{K}=\ket{3}$ at $\overline{s}=s(2,3)$. We therefore get requirements for our actual control $A$ by imposing $A(s(2,3)) \neq 0$ unlike for $\overline{A}$, for the rest copying the requirements of $\overline{A}$, i.e.~$A(s) = 0$ for $s \in \{s(0,3),s(1,3),s(0,2)\}$ and $A(s) \neq 0$ for $s \in \{s(0,1),s(1,2)\}$.
To satisfy these requirements, we take the polynomial control $A(s)=s(1-s)(s-s(0,3))(s-s(1,3))(s-s(0,2))$, represented on Fig.\ref{fig:fig3}.a.
Fig.\ref{fig:fig3}.b shows how the eigenvalues of $H(s)$ cross and anti-cross depending on whether $A(s)$ vanishes or not.
The squared norm components of $U^{\varepsilon}(1)$ resulting from a simulation of \eref{eq:dyn_U} with this control and $\varepsilon=10^{-3}$ are shown on Fig.\ref{fig:fig3}.d on a white-to-black scale, confirming achievement of permutation $\sigma(0,1,2,3)=(2,0,3,1)$.

\begin{figure}[!t]
\begin{center}
\setlength{\unitlength}{1mm}
\begin{picture}(140,64)(0,15)
	\put(0,64){\includegraphics[width=20mm]{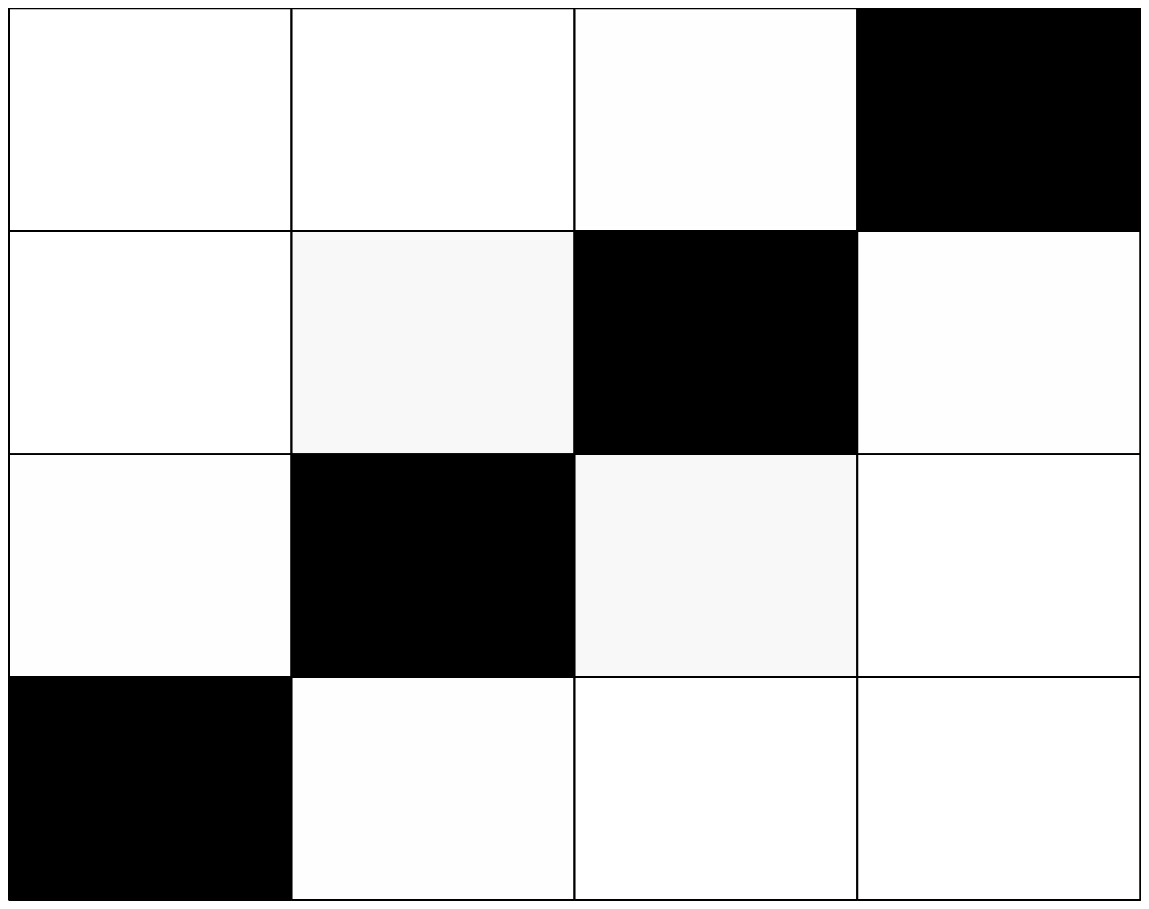}}
	\put(20,64){\includegraphics[width=20mm]{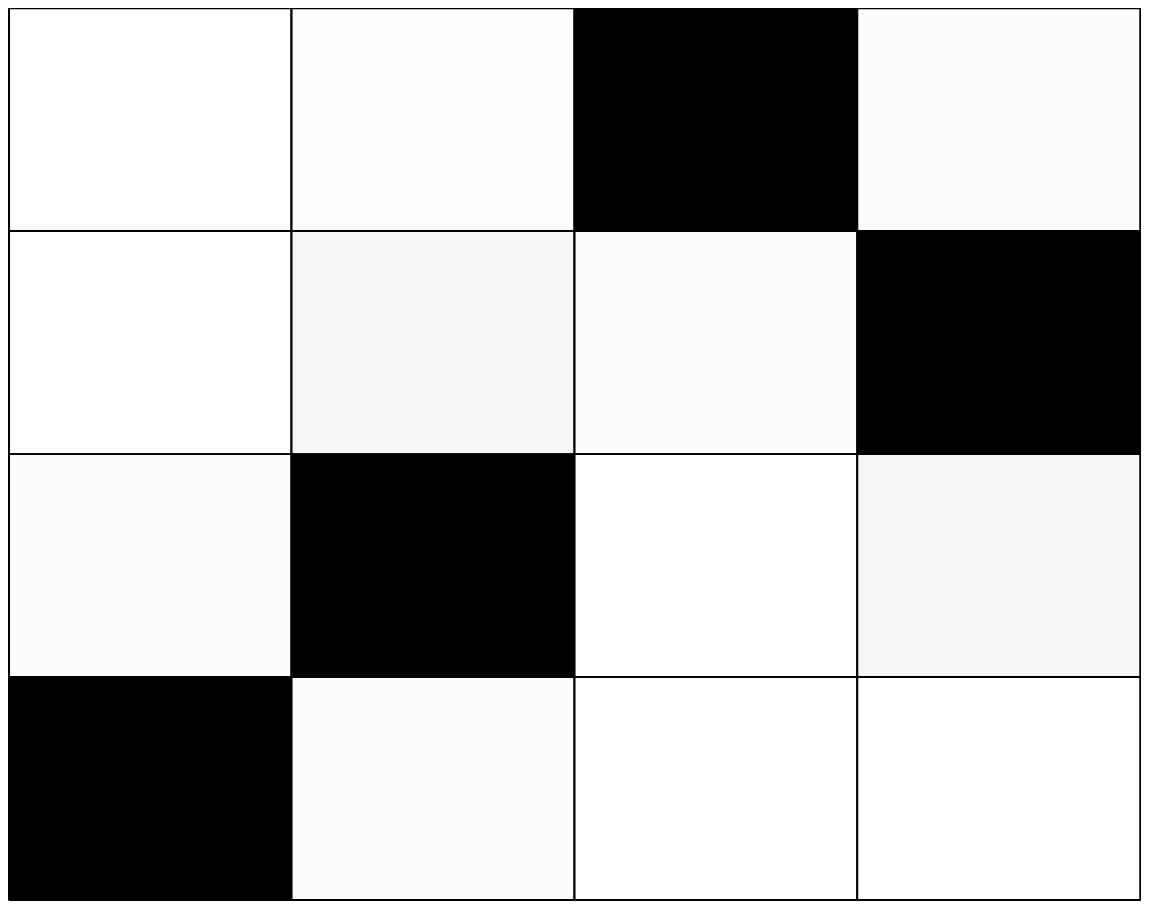}}
	\put(40,64){\includegraphics[width=20mm]{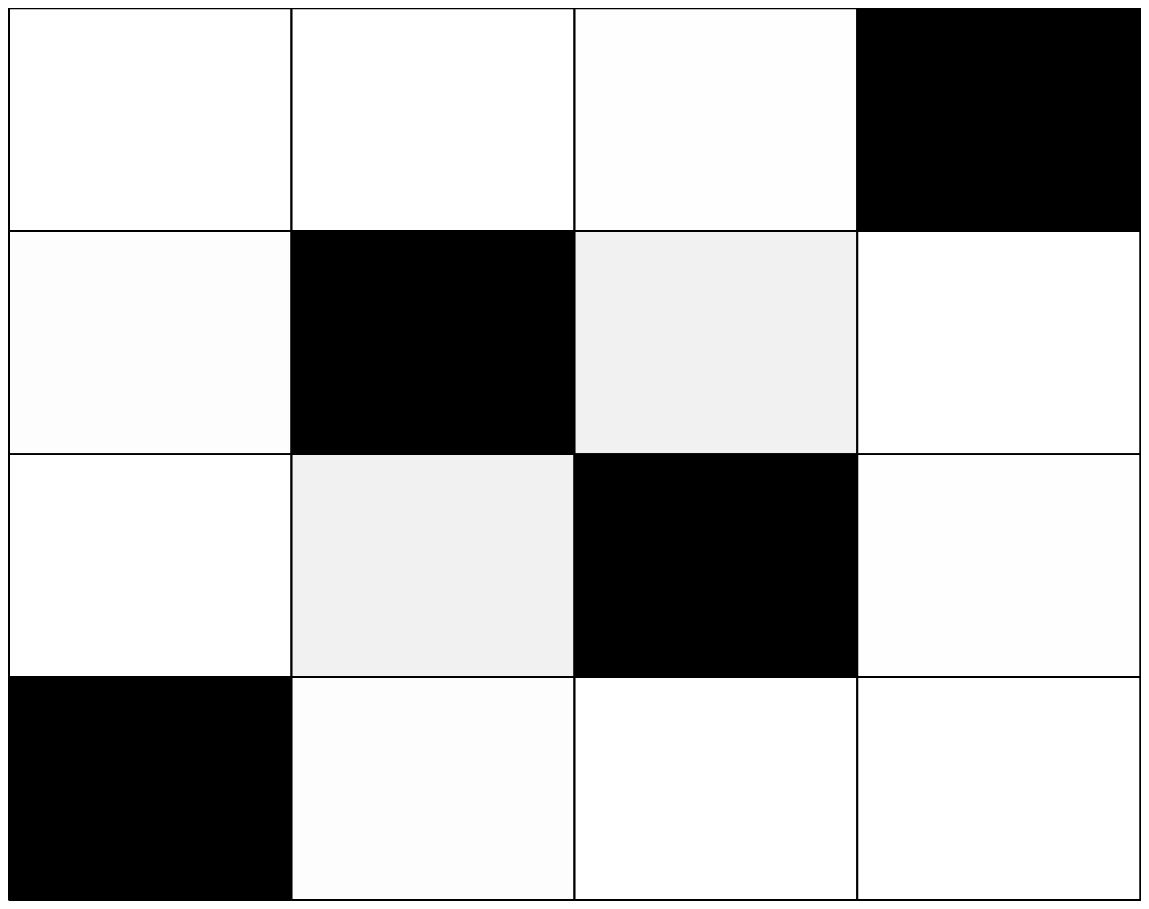}}
	\put(60,64){\includegraphics[width=20mm]{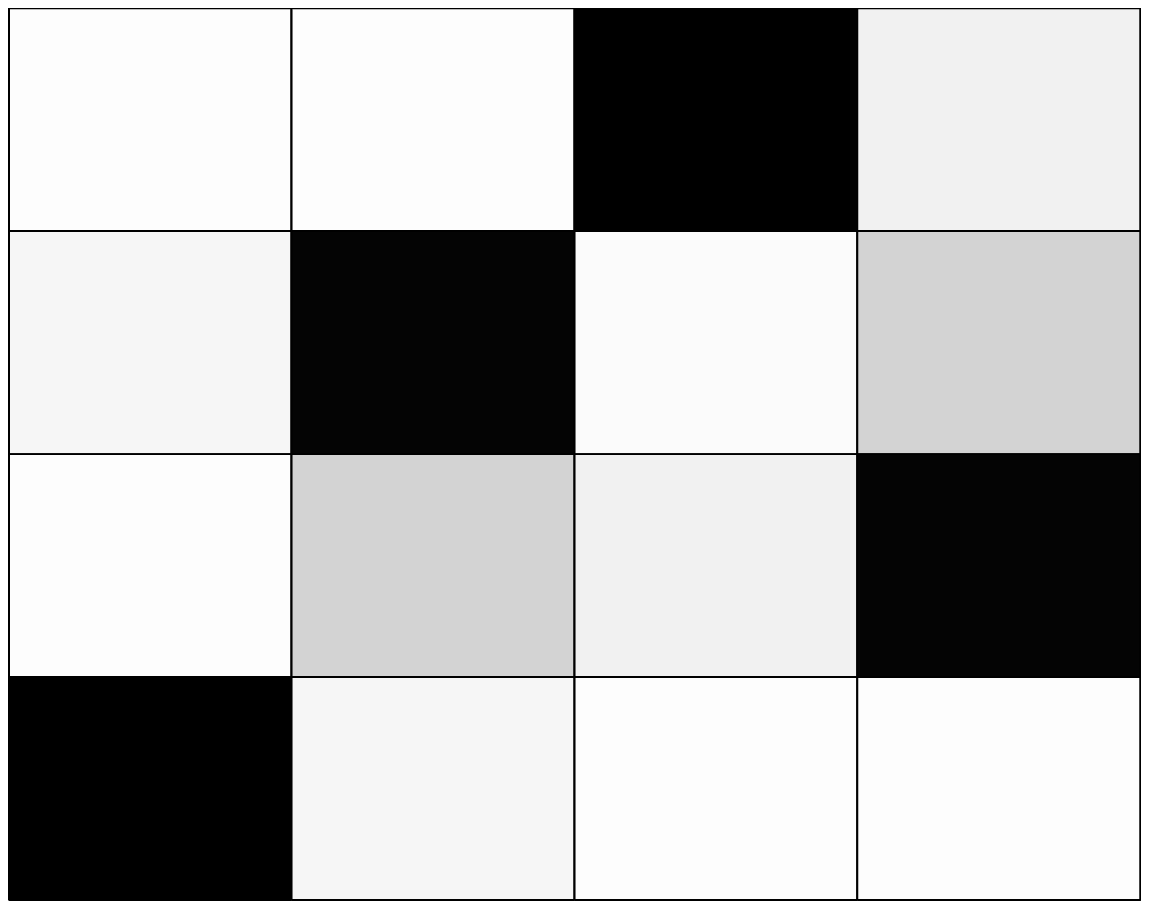}}
	\put(80,64){\includegraphics[width=20mm]{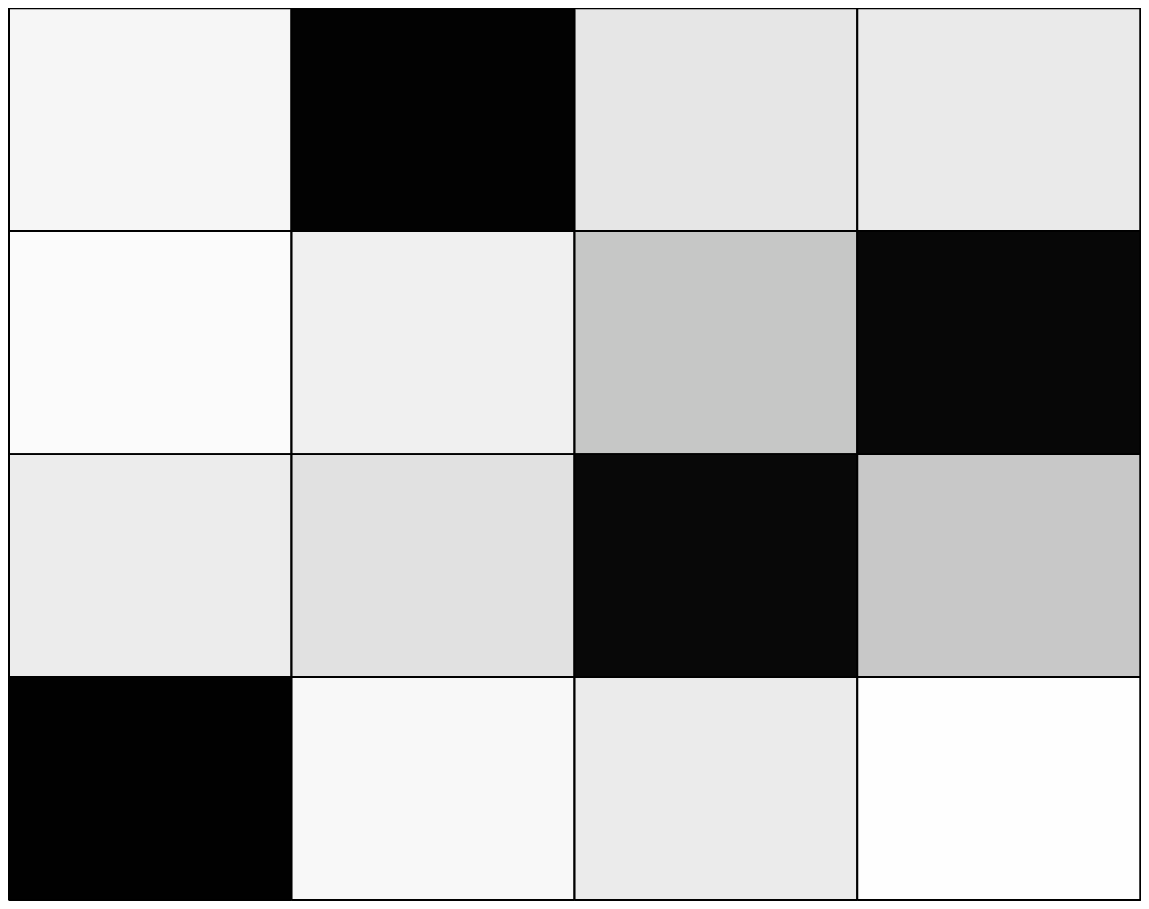}}
	\put(100,64){\includegraphics[width=20mm]{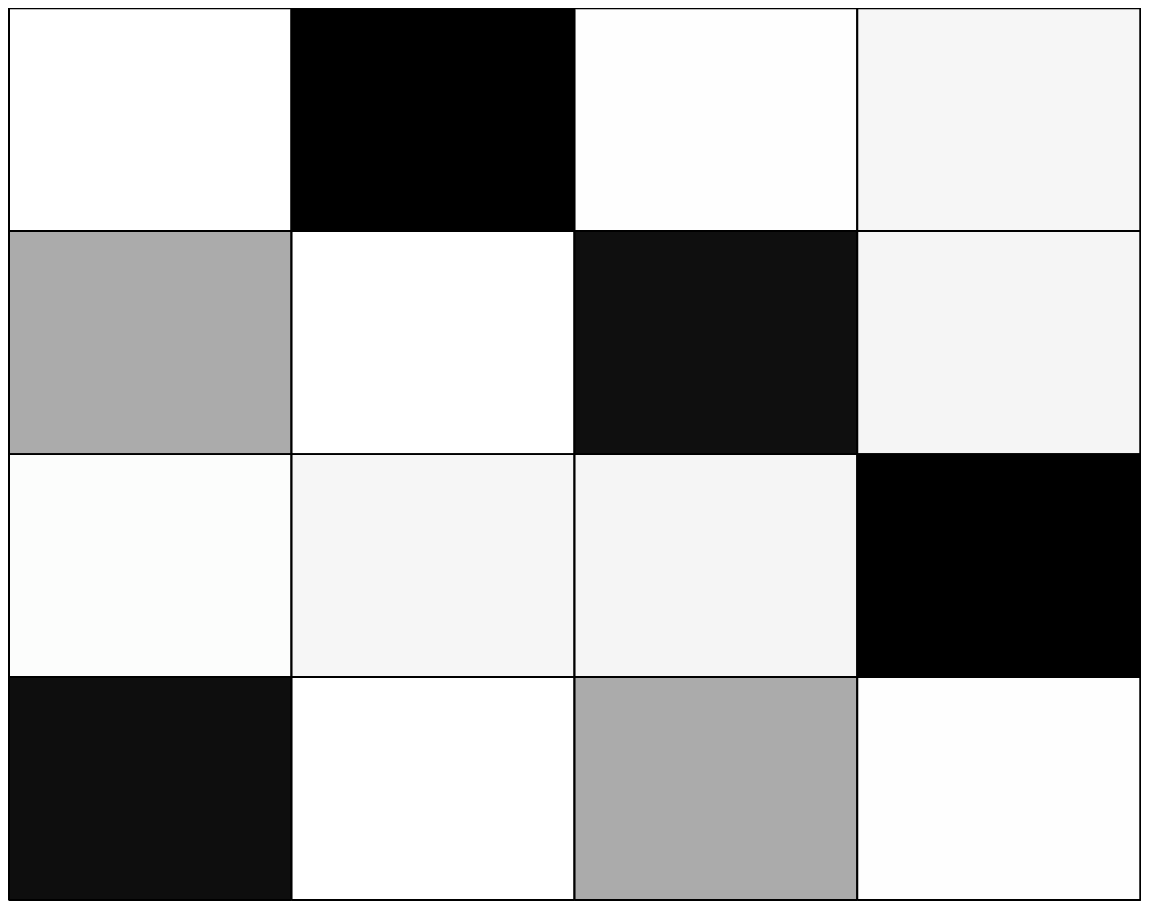}}
	\put(120,64){\includegraphics[width=20mm]{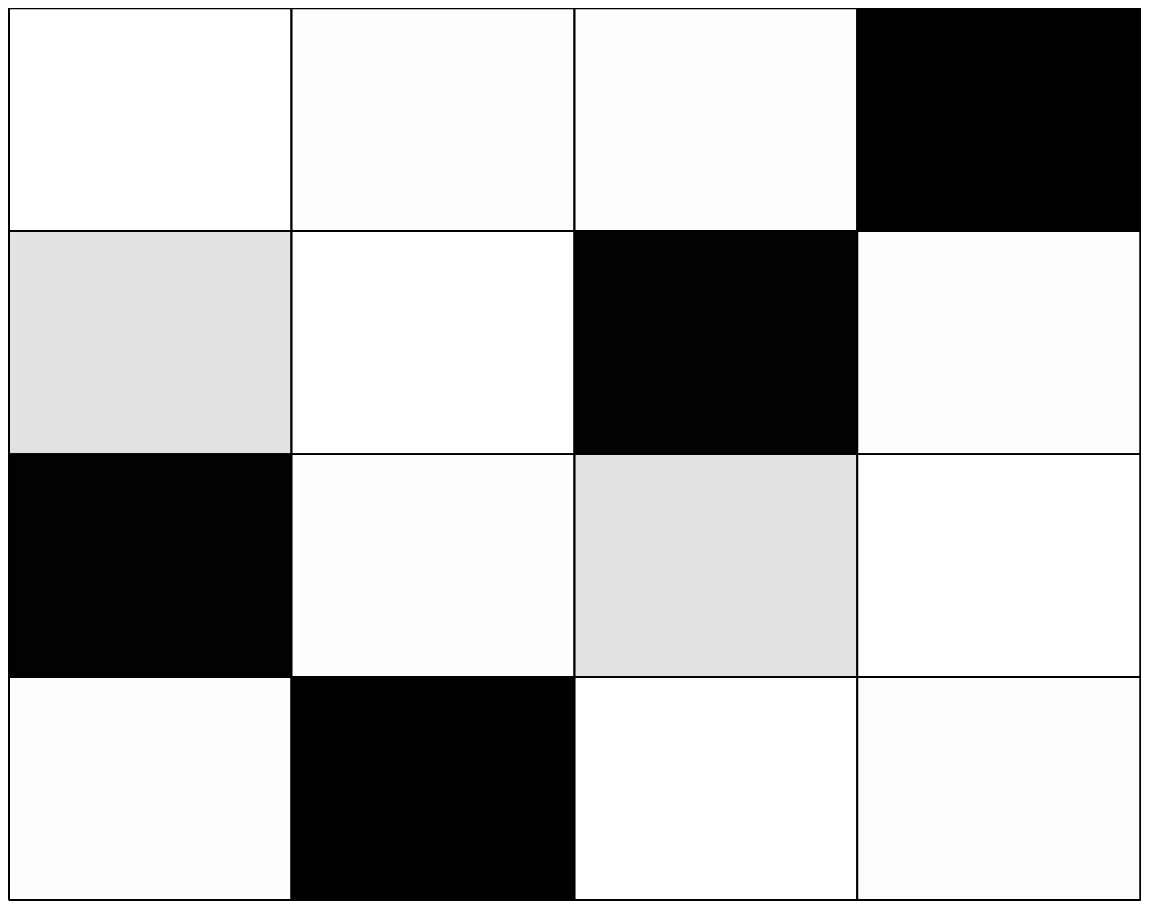}}
	\put(0,16){\includegraphics[width=20mm]{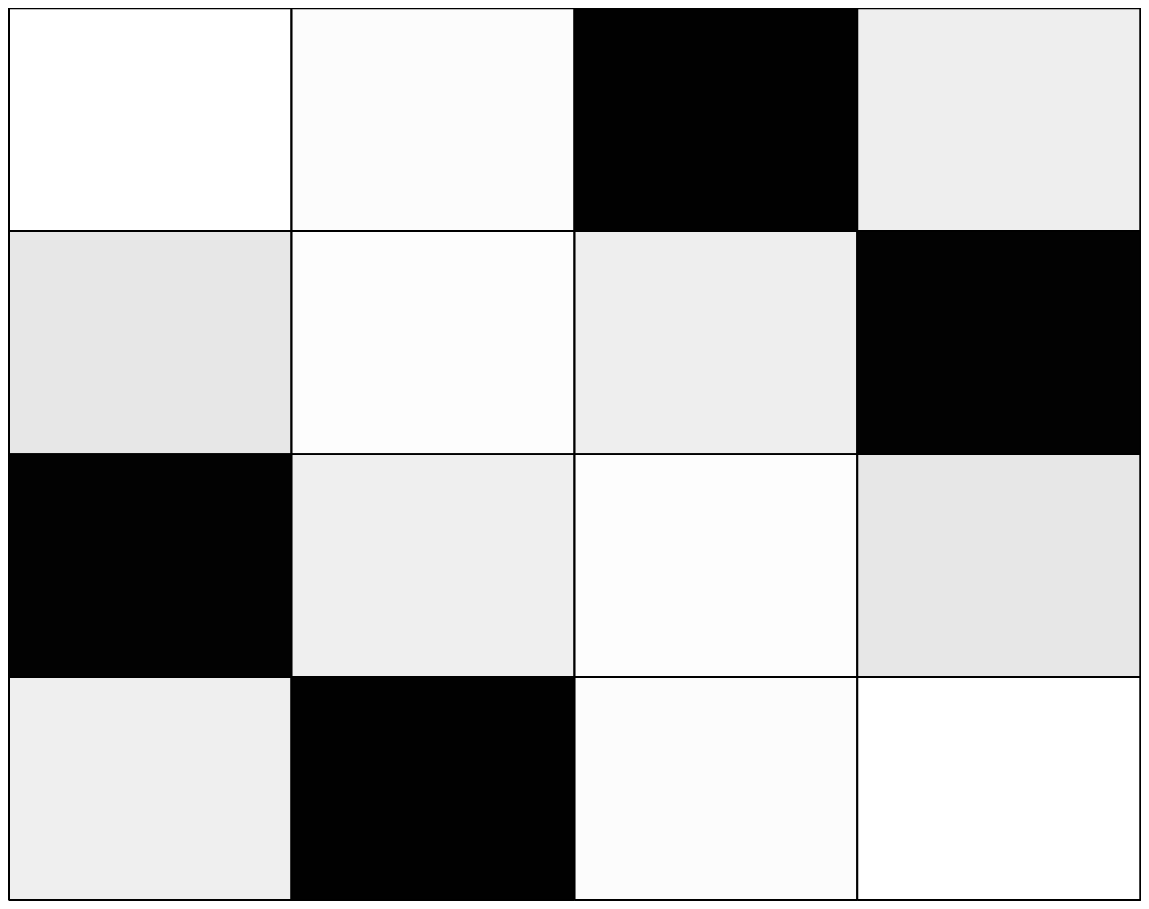}}
	\put(0,48){\includegraphics[width=20mm]{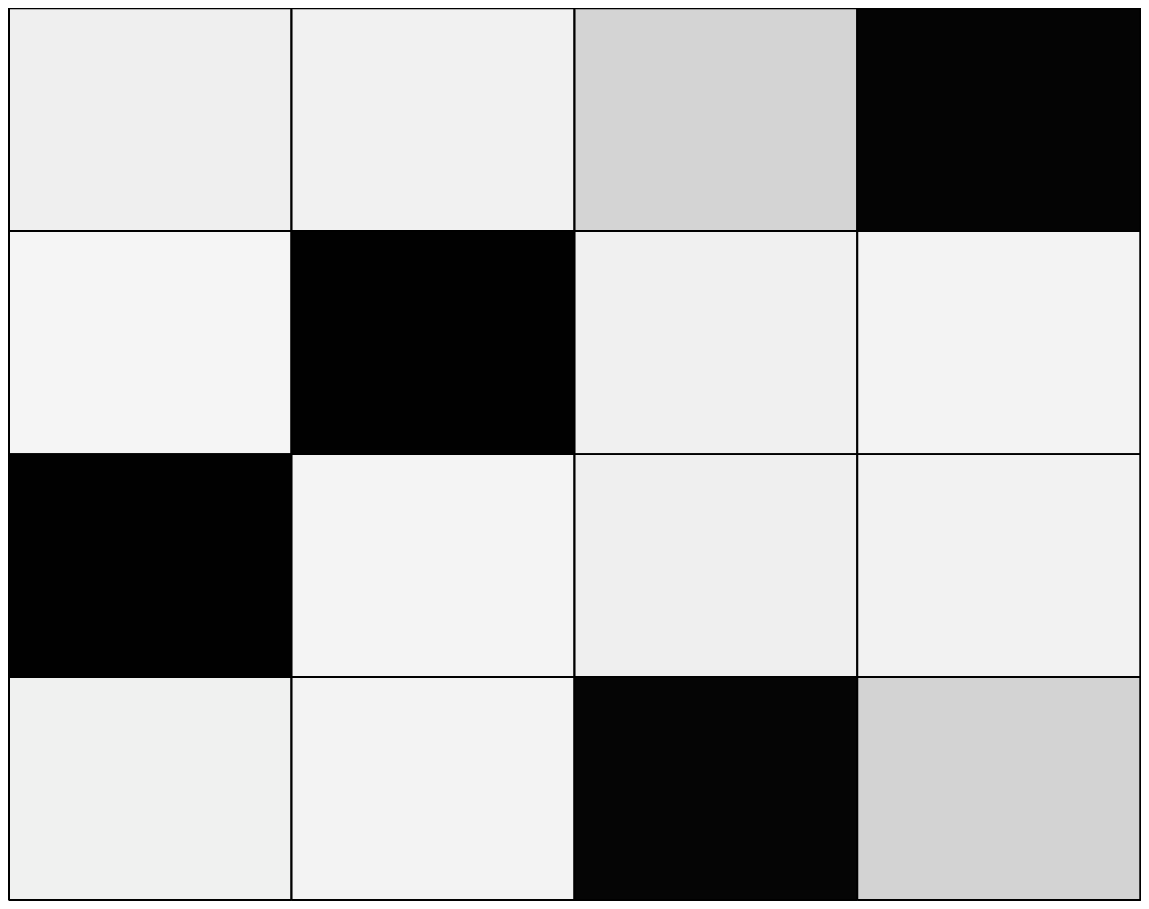}}
	\put(20,48){\includegraphics[width=20mm]{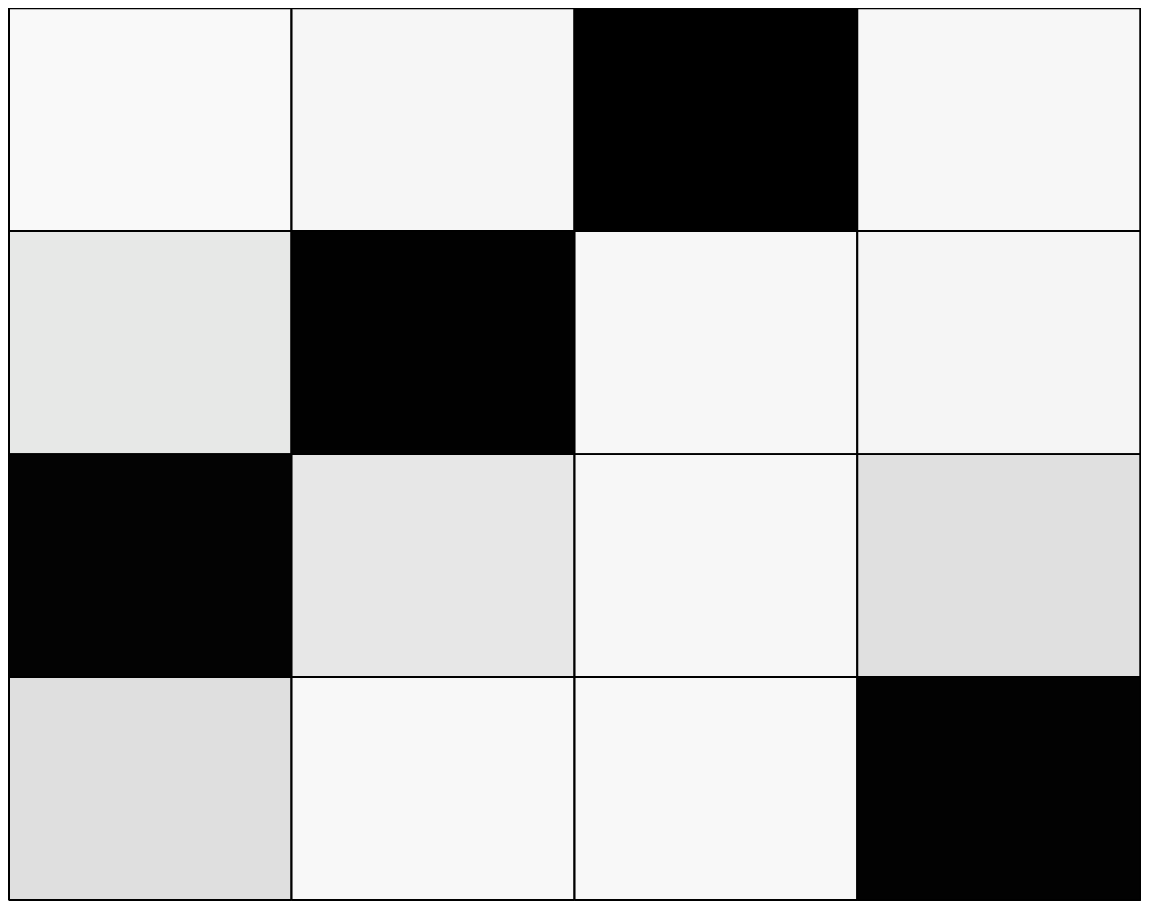}}
	\put(40,48){\includegraphics[width=20mm]{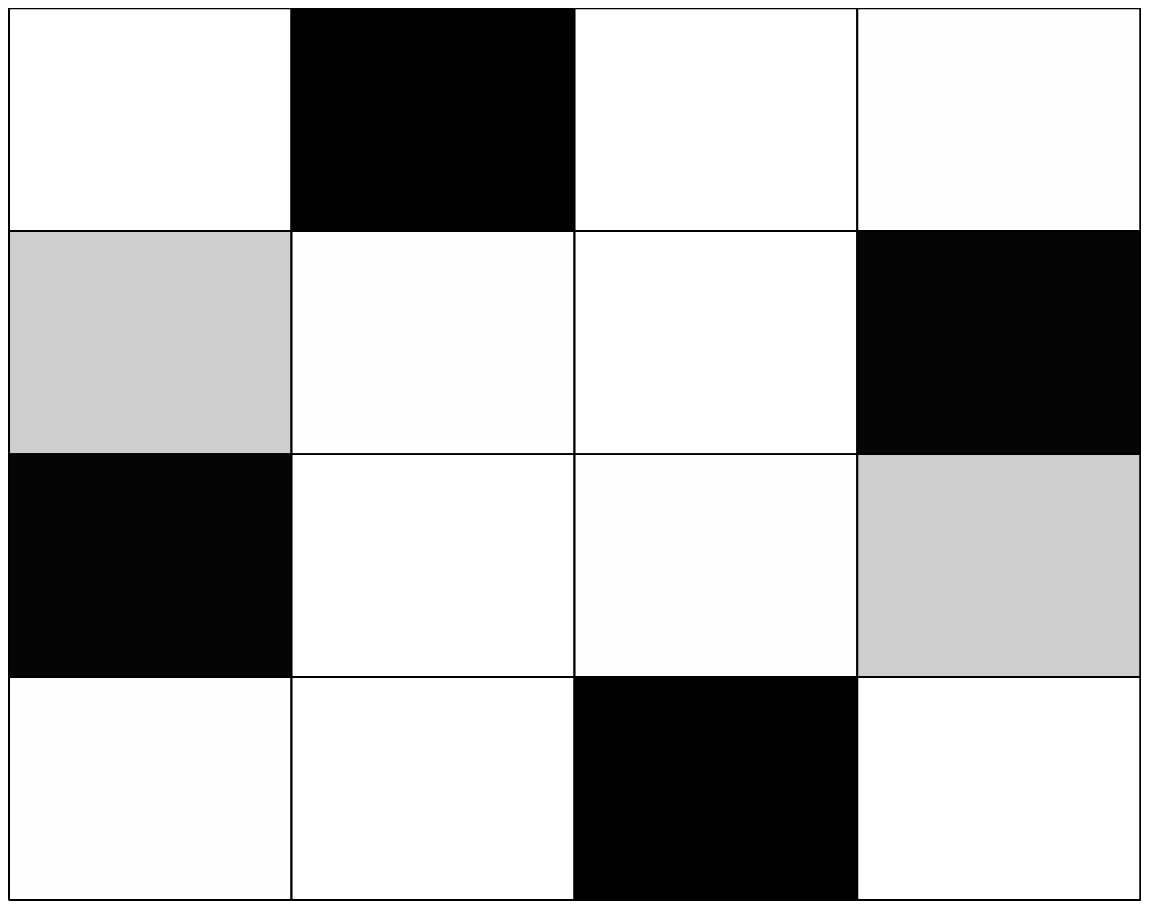}}
	\put(60,48){\includegraphics[width=20mm]{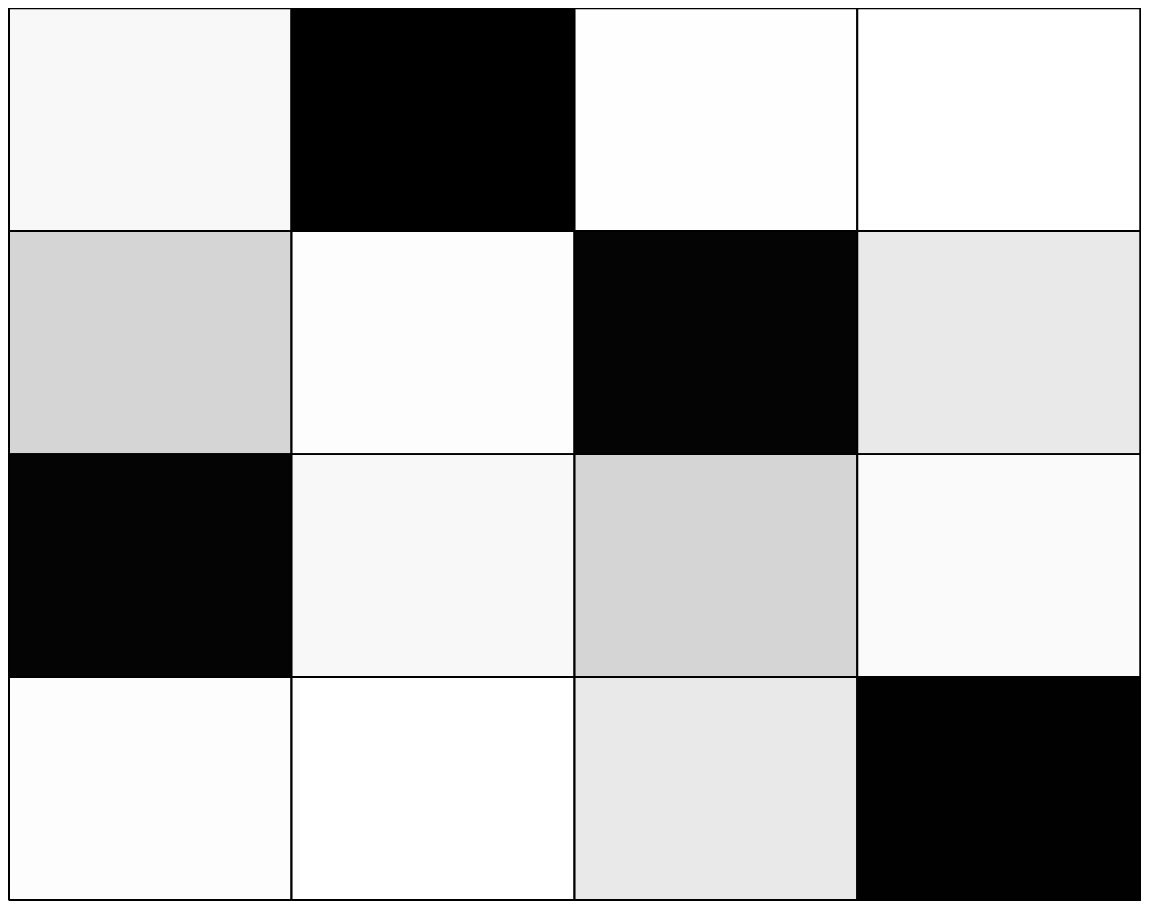}}
	\put(80,48){\includegraphics[width=20mm]{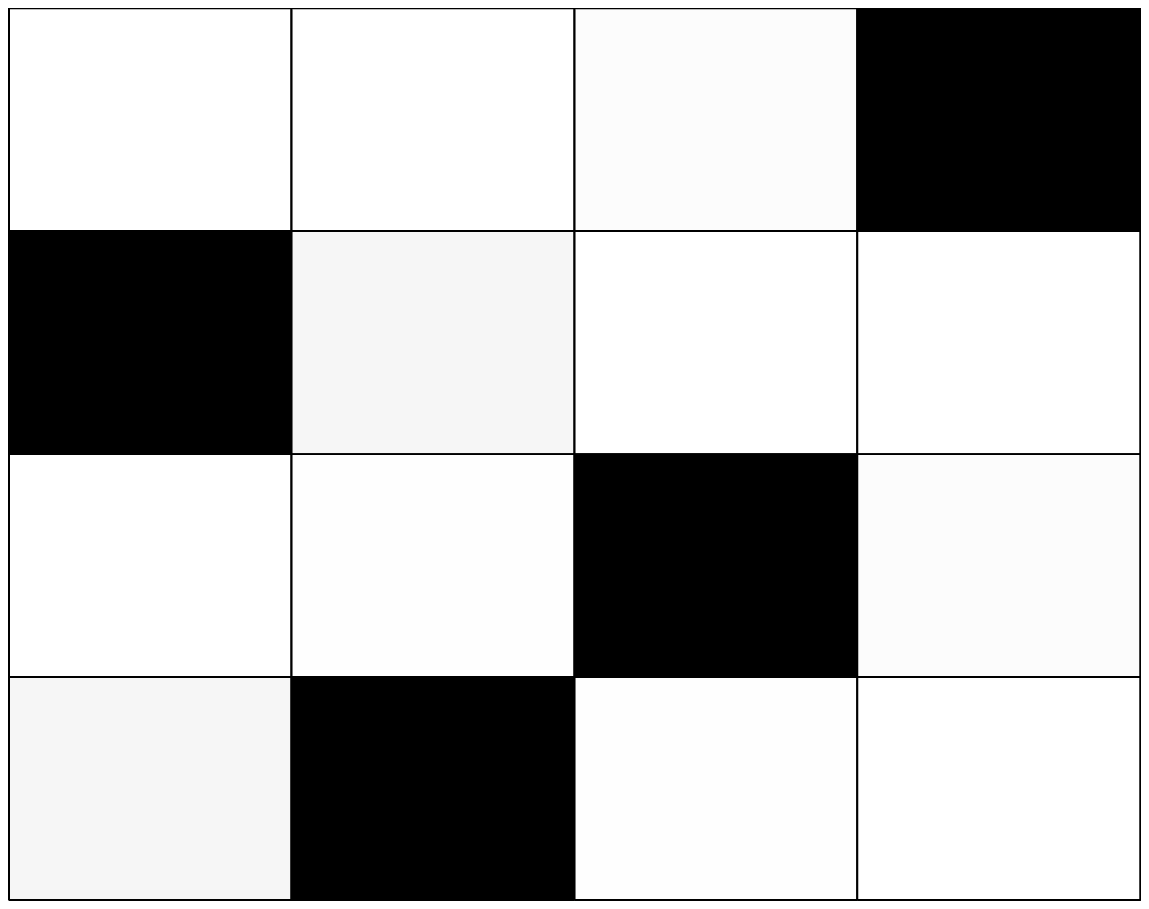}}
	\put(100,48){\includegraphics[width=20mm]{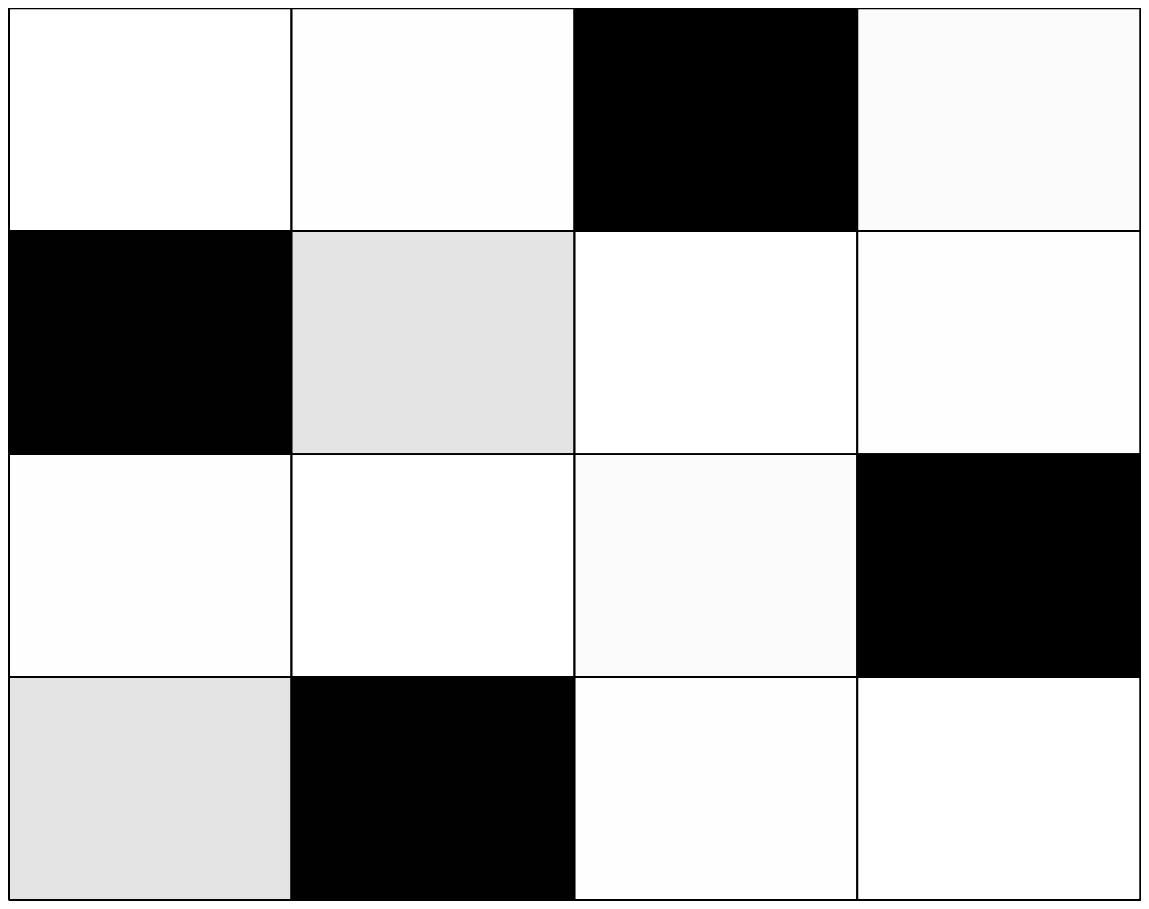}}
	\put(120,48){\includegraphics[width=20mm]{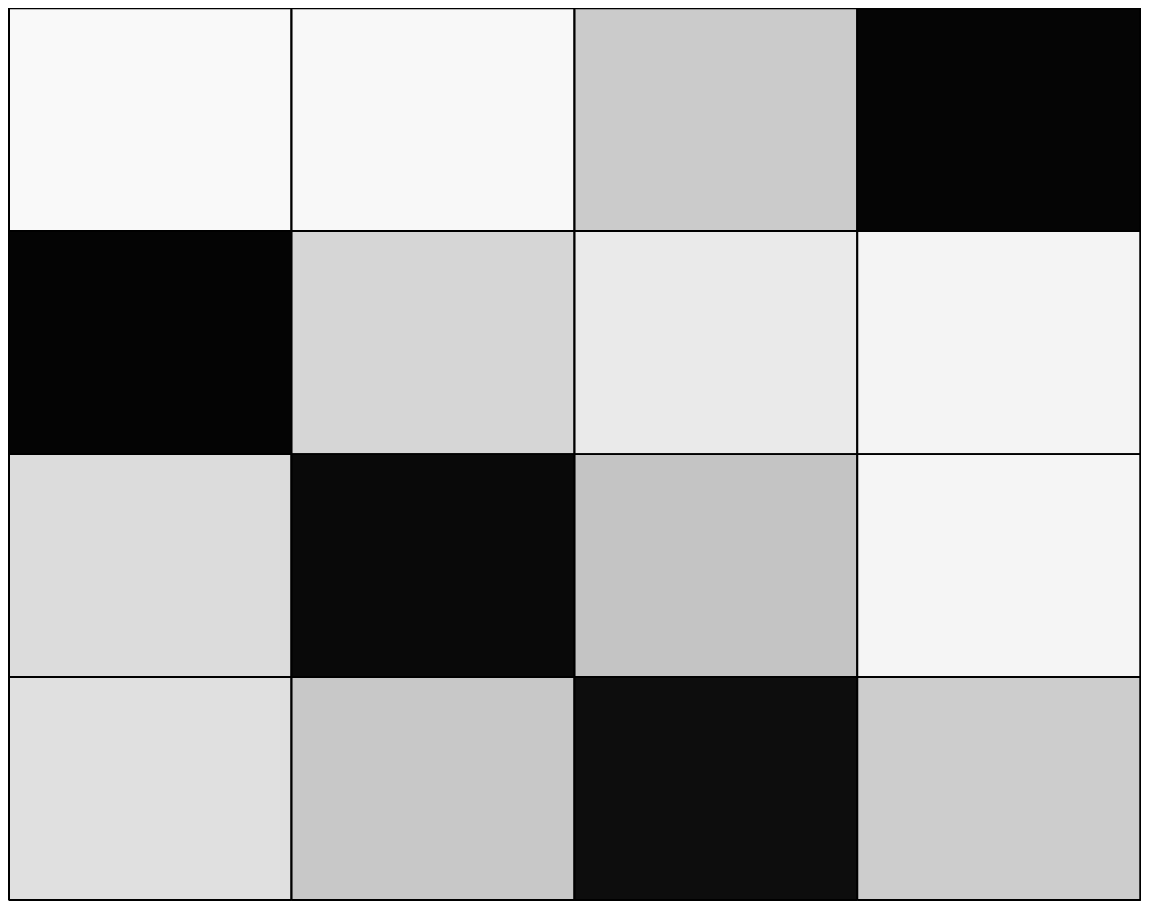}}
	\put(20,16){\includegraphics[width=20mm]{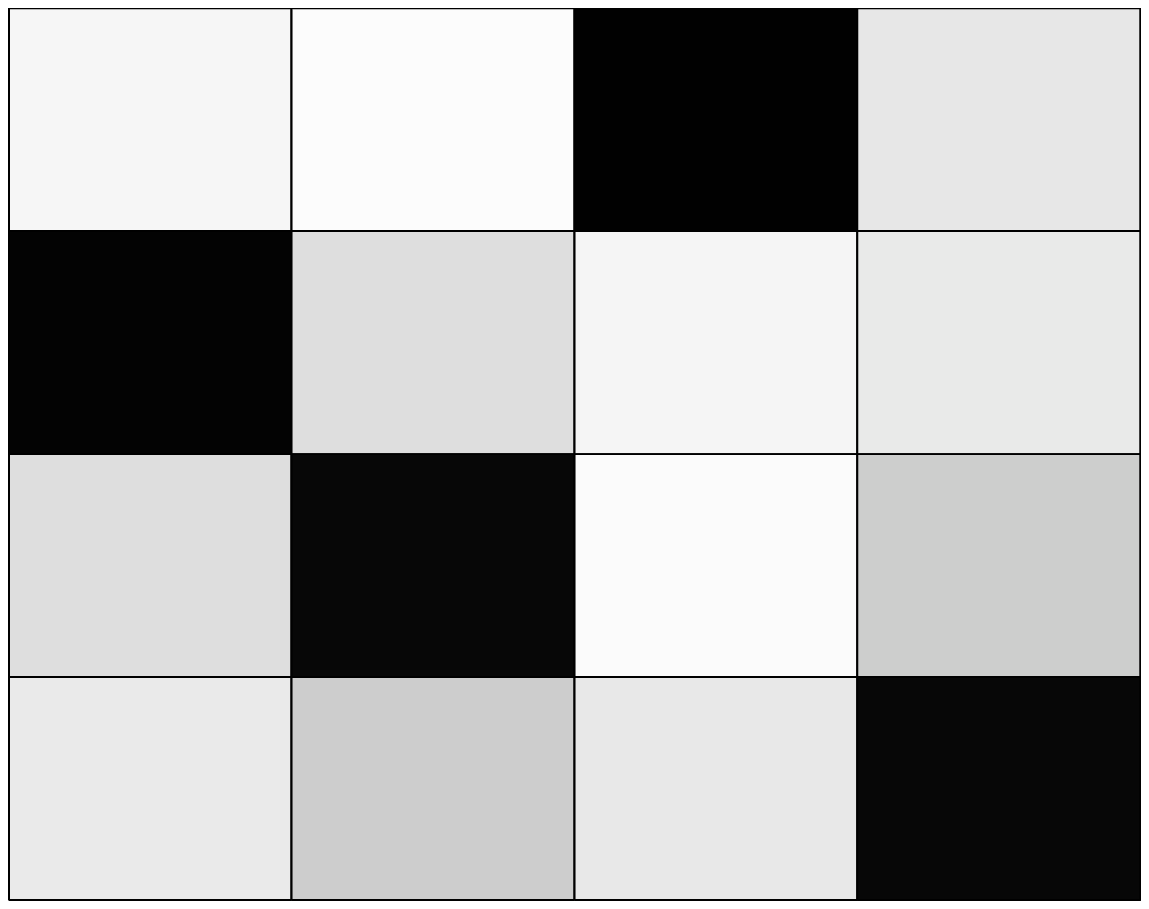}}
	\put(0,32){\includegraphics[width=20mm]{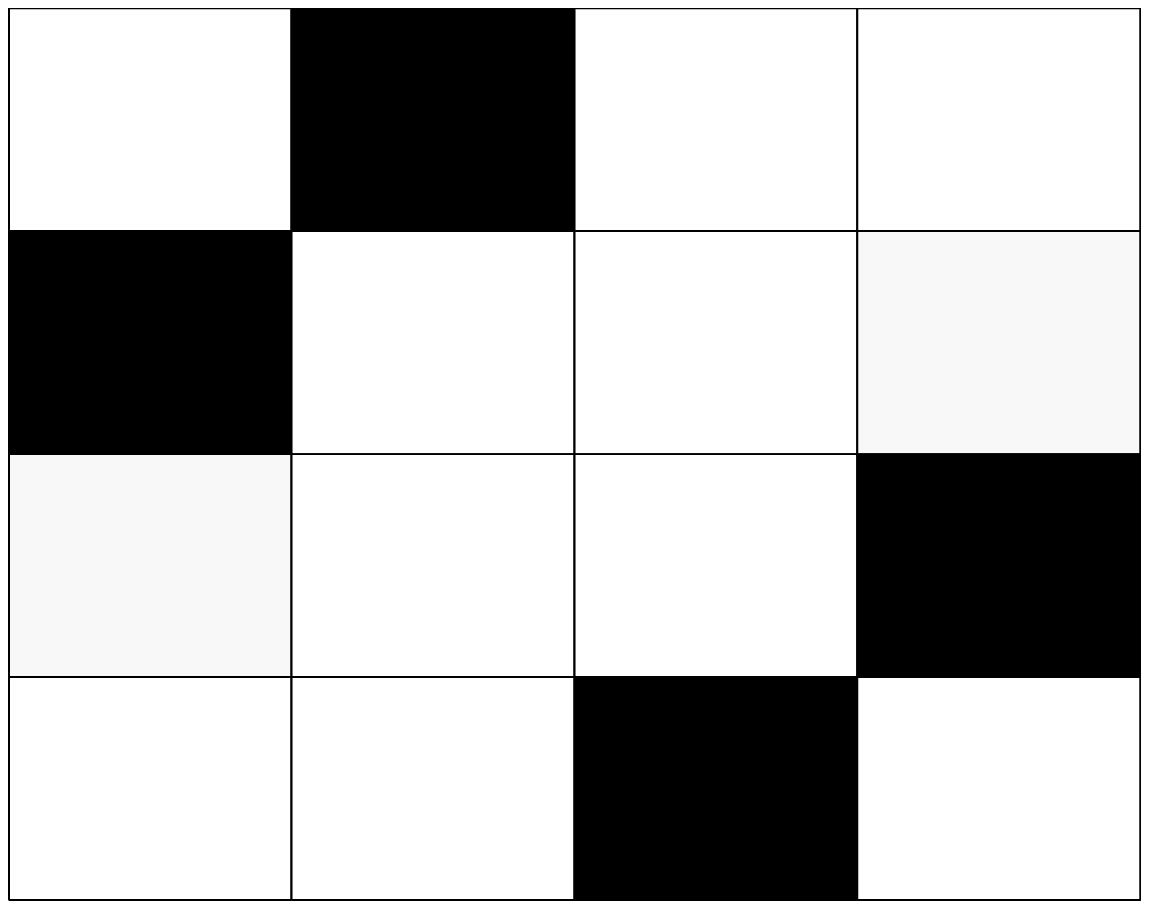}}
	\put(20,32){\includegraphics[width=20mm]{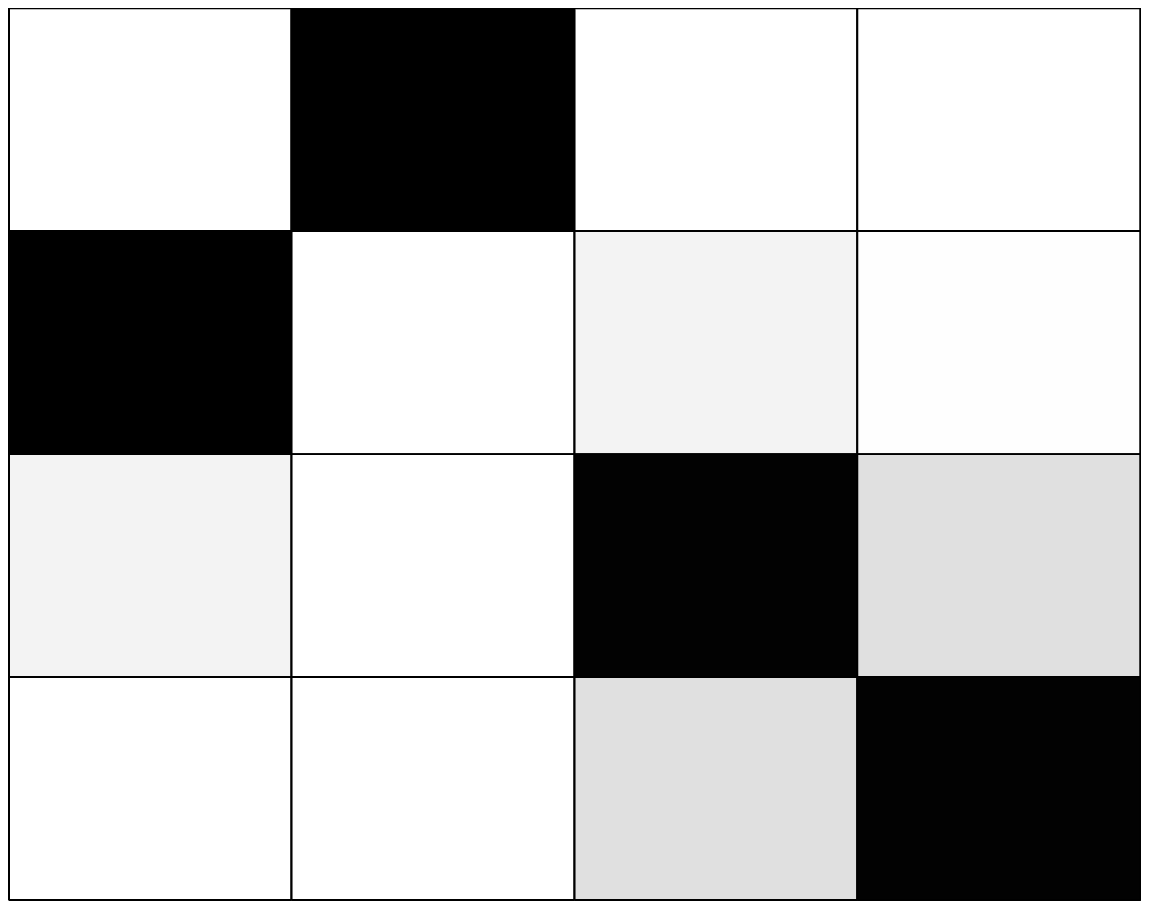}}
	\put(40,32){\includegraphics[width=20mm]{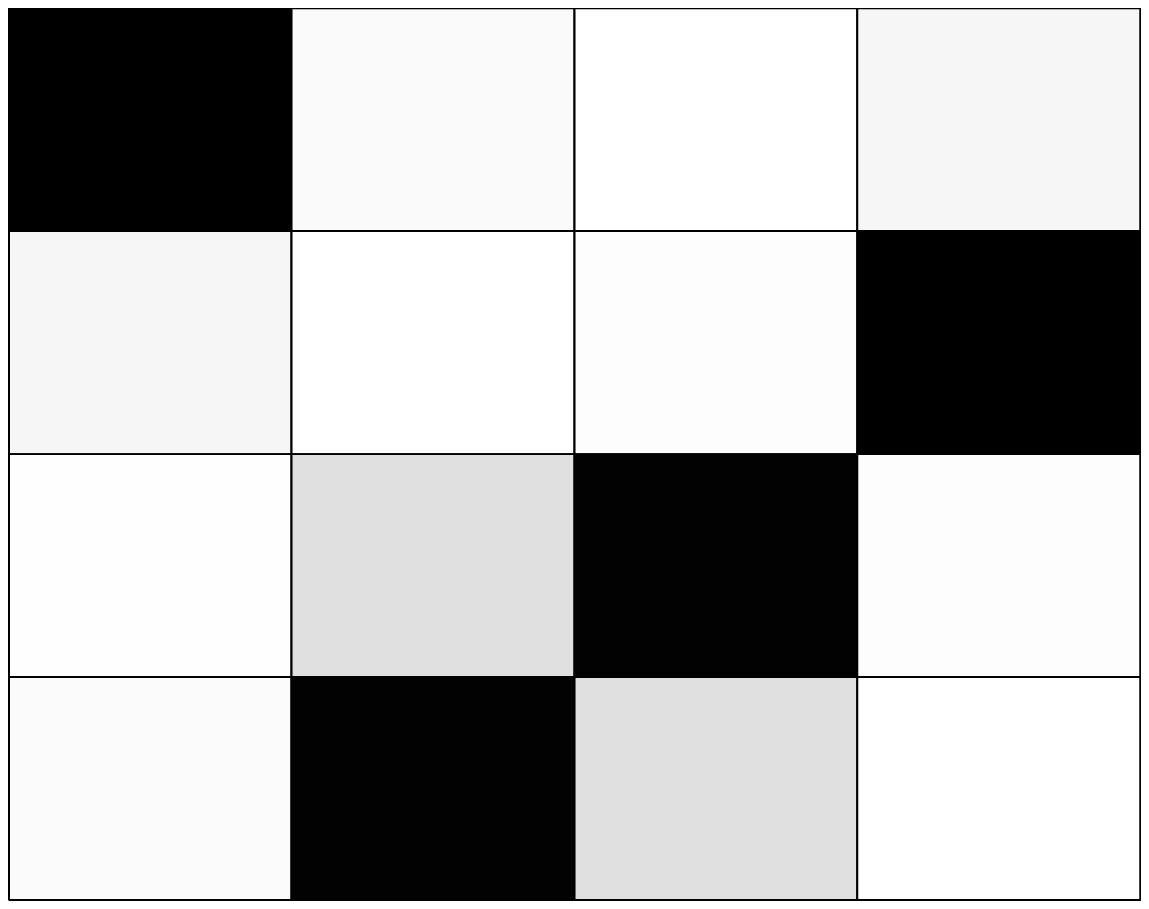}}
	\put(60,32){\includegraphics[width=20mm]{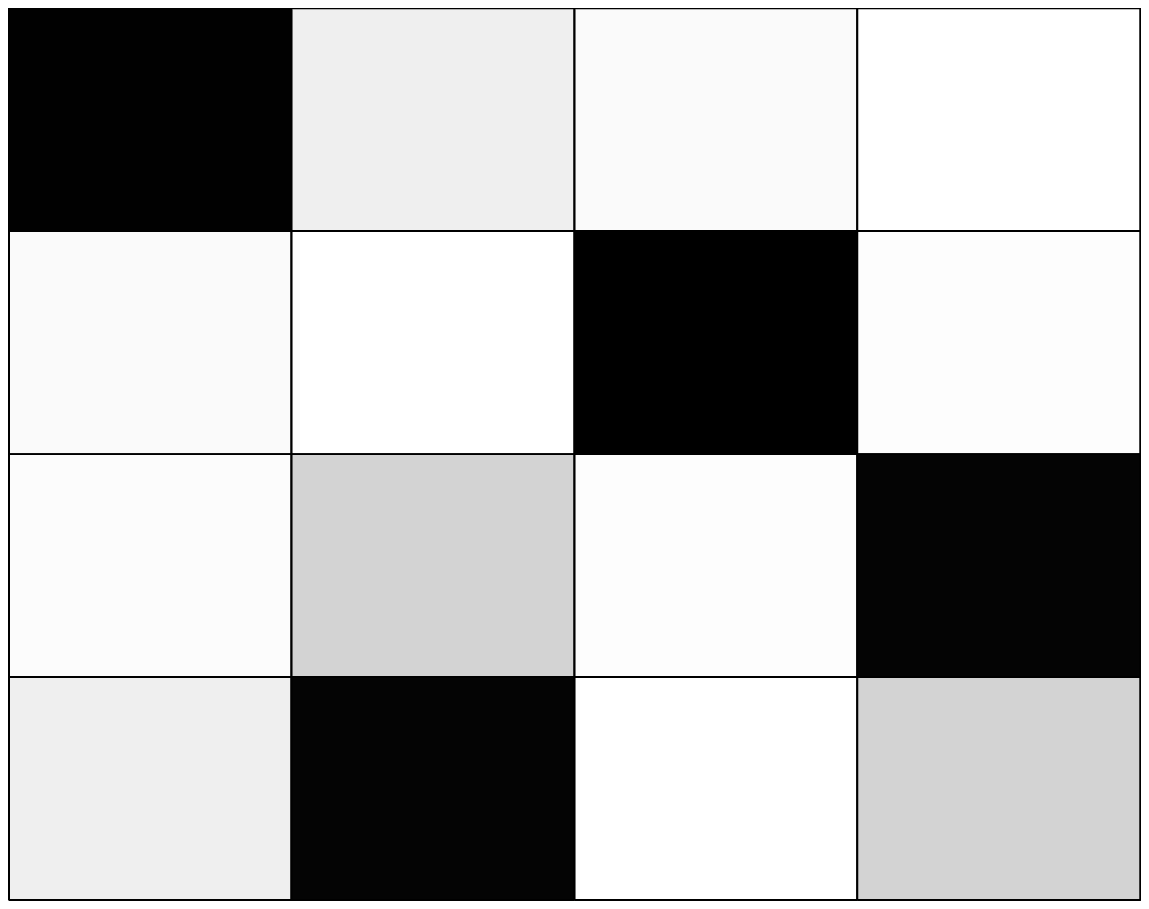}}
	\put(80,32){\includegraphics[width=20mm]{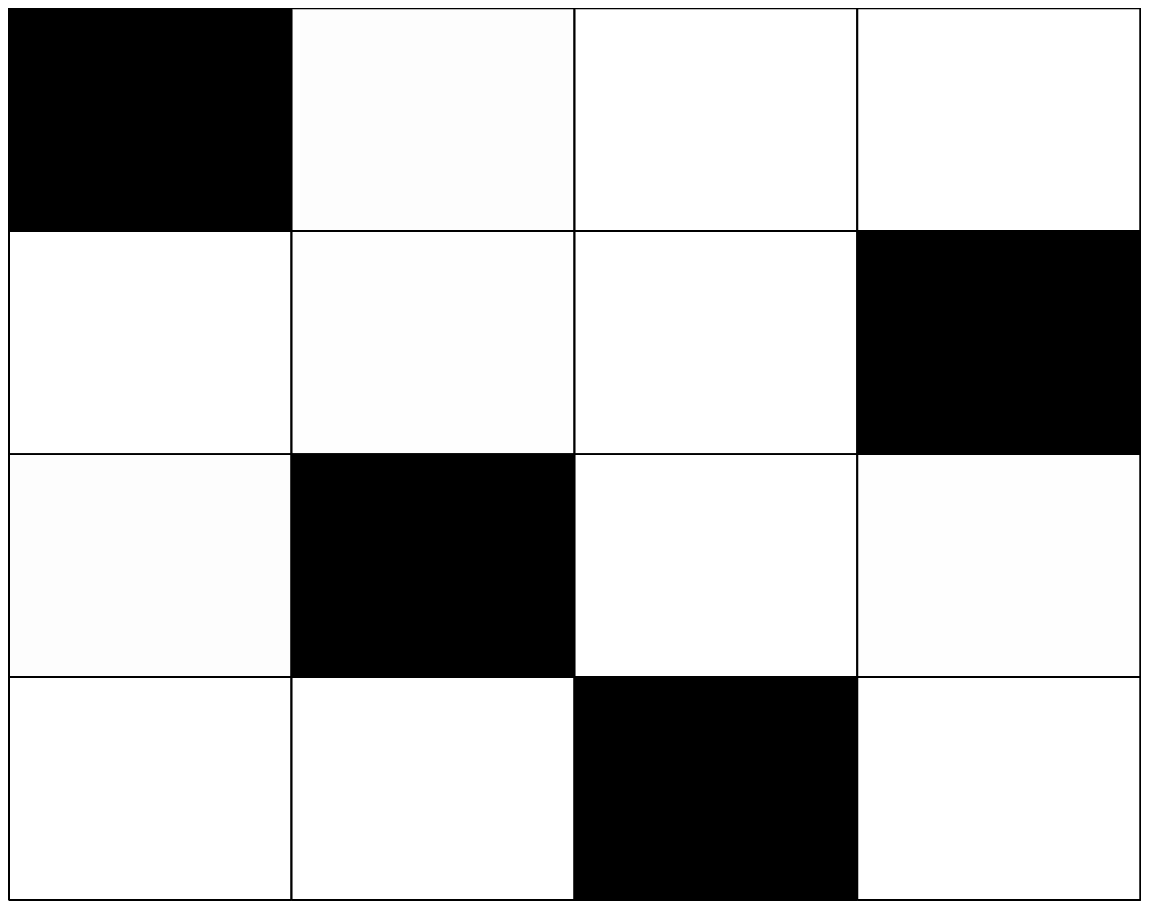}}
	\put(100,32){\includegraphics[width=20mm]{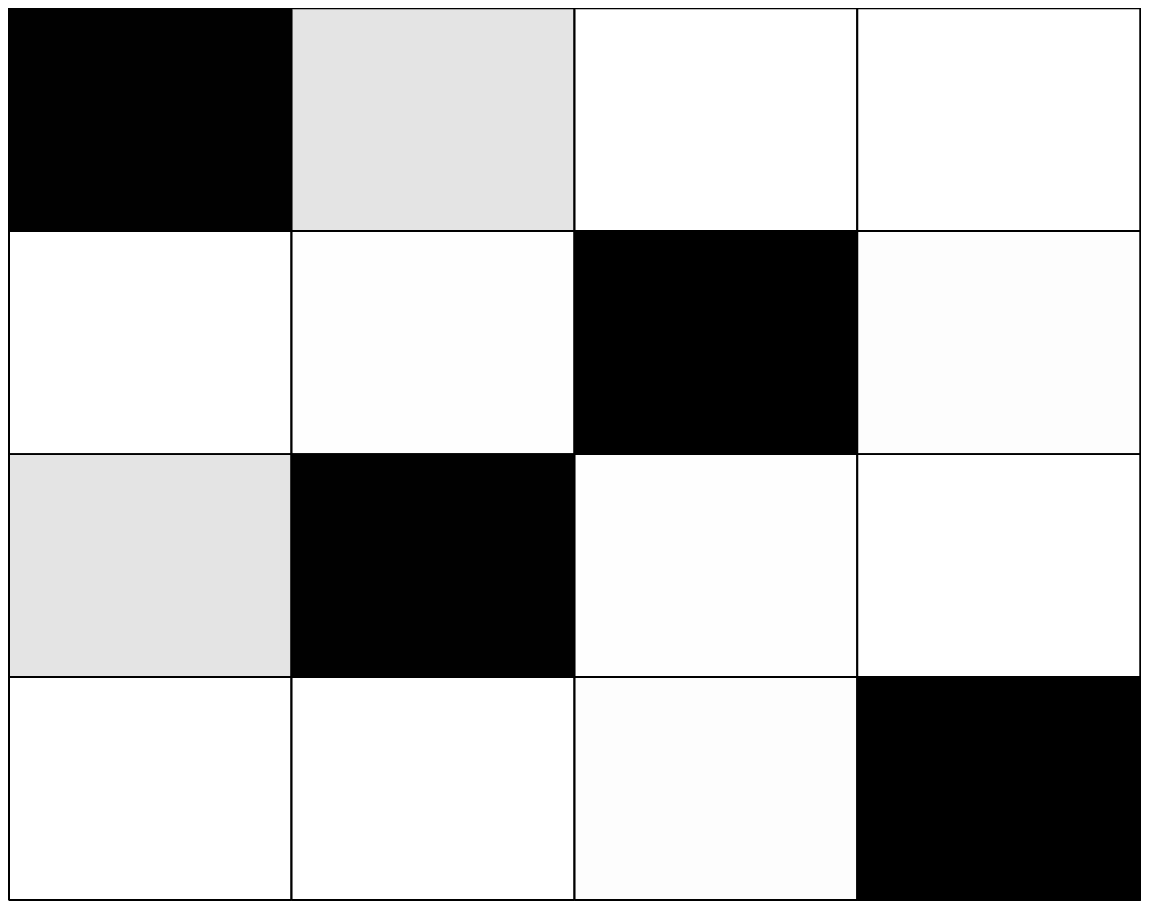}}
	\put(120,32){\includegraphics[width=20mm]{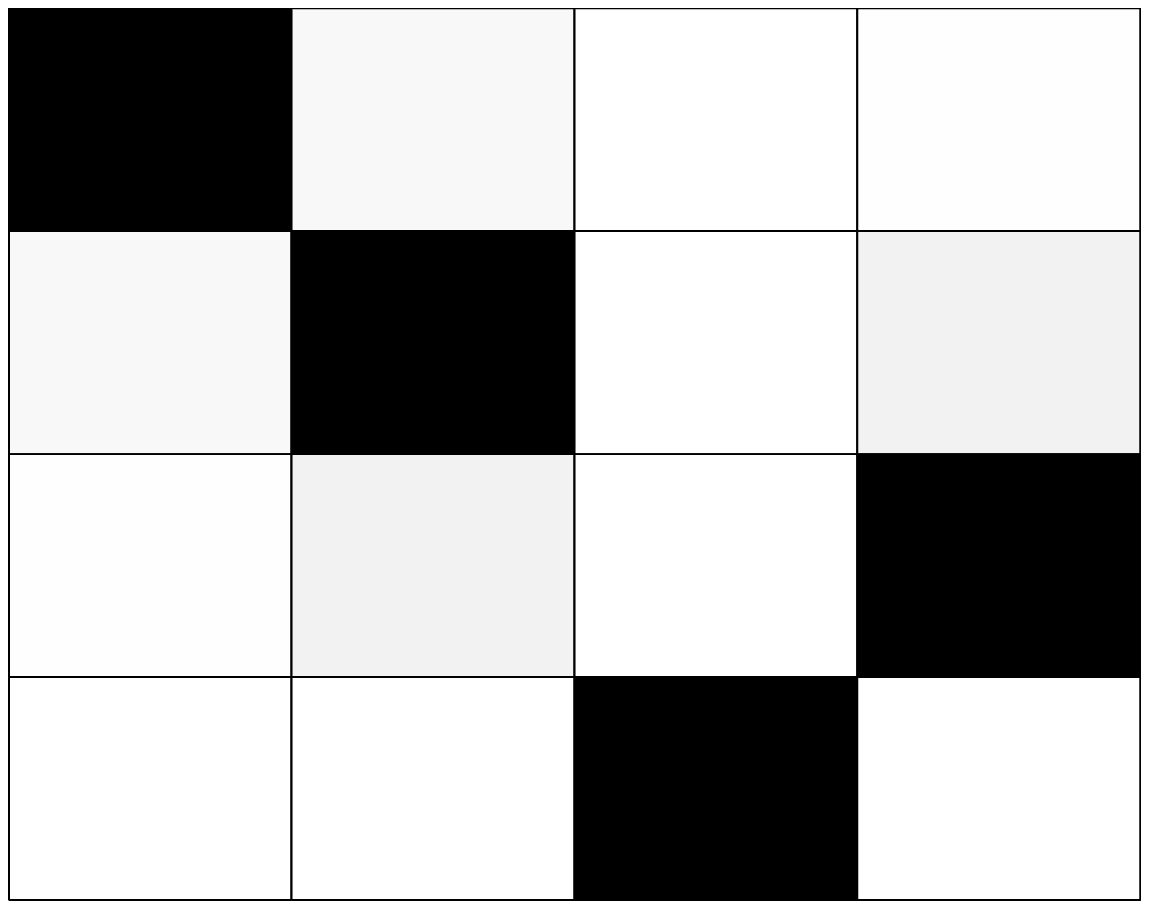}}
	\put(40,16){\includegraphics[width=20mm]{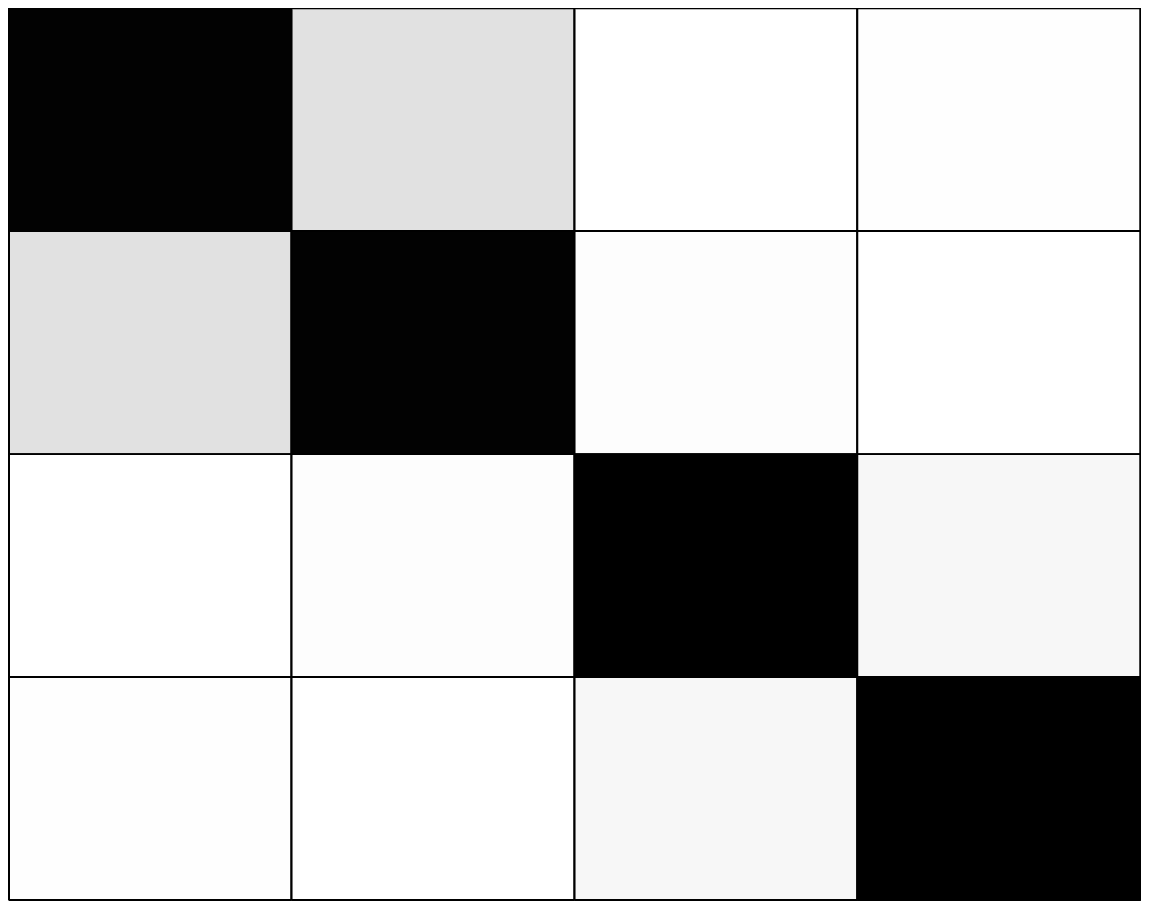}}
	\put(61,16){\includegraphics[height=16mm]{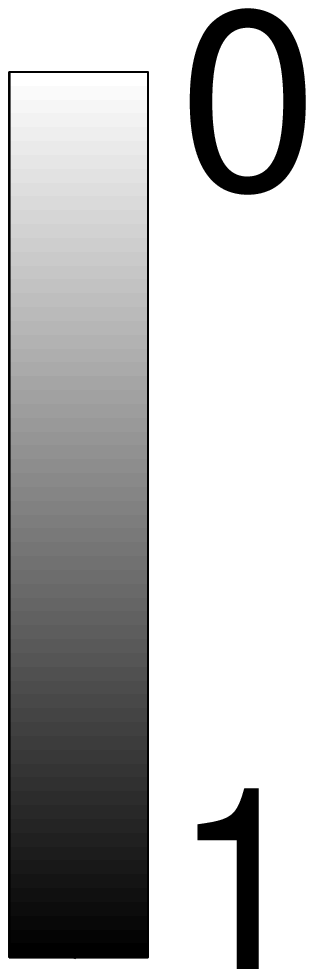}}	
\end{picture}
\caption{Simulation results of \eref{eq:dyn_U} for 24 different controls $A(s)$ following Theorem \ref{thm:3} to achieve each one of the 24 permutations of $(0,1,2,3)$ with adiabatic passage. Shading represents squared norm of elements of matrix $U^{\varepsilon}(1)$ expressed in basis $\ket{0},\ldots,\ket{3}$, from white (value 0) to black (value 1). In other words, each subplot may be read as a $4 \times 4$ matrix where the black patches are ones and the white patches are zeros; gray patches indicate intermediate values, reflecting that the unitary propagator obtained by integrating \eref{eq:dyn_U} is not exactly a permutation matrix for the finite $\varepsilon=10^{-3}$.}
\label{fig:allperms}
\end{center}
\end{figure}

Fig.\ref{fig:allperms} shows the same squared norm components of $U^{\varepsilon}(1)$ in gray-shades for 24 cases, corresponding to
different
control inputs $A(s)$ designed for all $24$ possible permutations of the set $(0,1,2,3)$. The controls $A(s)$ are built as the
product of (i) a polynomial
vanishing on $\mathcal{I}_A \cup \{ 0,1\}$ and only there,
and (ii) a set of functions $(1+g(s-s(m,n)))$, with $g(s-s(m,n))$ Gaussians centered on all
$s \in \mathcal{I}_\omega \setminus \mathcal{I}_A$;
the role of the latter is to amplify $A(s)$ in the vicinity of intended ``anti-crossings'', improving convergence of the adiabatic limit as a function of $\varepsilon$. Fig.\ref{fig:allperms} corresponds to the choice $\varepsilon=10^{-3}$.

\section{Proofs}\label{sec:proofs}

In this section we give the proofs of all the formal results presented in previous sections.


\subsection{Proof of Theorem \ref{thm:1}}
We start the proof by recalling the following result \cite{Nagao-Wadati_JPSJ_1993}.
\begin{lm}
\label{Lemma:nondegeneracy}
Let $D_N$ be a real tridiagonal and symmetric $N\times N$ matrix defined by
\begin{equation}\label{eq:DN}
D_N=\sum_{k=0}^{N-1}a_k\proj{k}+\sum_{k=0}^{N-2}c_k(\ket{k}\bra{k+1}+\ket{k+1}\bra{k})
\end{equation}
in some orthonormal basis $(\ket{0},\ldots,\ket{N-1})$.
If $c_k\ne 0$ for all $k\in \NN_0^{N\minou 2}$, then $D_N$ is non degenerate.
\end{lm}
\begin{pf}[of Lemma \ref{Lemma:nondegeneracy}]
Denote $Q_n$ the characteristic polynomial of $D_n$, which is defined
as \eref{eq:DN} with $N$ replaced by $n$,
for all $n\in\{1,\ldots,N\}$. The sequence of polynomials $(Q_n)_n$ verifies the following recurrence relation: for $n\ge 2$,
\begin{equation}
\label{eq:recurrence}
Q_n(x)=(x-a_{n-1})Q_{n-1}(x)-\left(c_{n-2}\right)^2Q_{n-2}(x)\, ,
\end{equation}
with $Q_0(x)=1$ and $Q_1(x)=x-a_0$. According to Favard's Theorem \cite{Favard-AcadScParis_1935}, a sequence verifying \eref{eq:recurrence} where $\left(c_{n-2}\right)^2>0$ for all $n$, is a sequence of orthogonal polynomials.
Furthermore, from \cite[Theorem 3.3.1]{Szego-AMS_1967}, every polynomial $Q_n$ in a sequence of orthogonal polynomials has $n$ real and distinct zeros; this is in particular true for $n=N$, therefore $D_N$ is non degenerate.
\hfill $\boldsymbol{\Box}$
\end{pf}

\begin{pf}[of Theorem \ref{thm:1}]
We prove the result for any single system in $\SSS$ and conclude that it remains true for the sup over $\SSS$.
Indeed, the application
\begin{eqnarray*}
\lefteqn{(\mu_0,\ldots,\mu_{N-2},\Delta_0,\ldots,\Delta_{N-1})} & & \\
& \phantom{l} & \rightarrow \;\;\; \Vert \; U^\varepsilon(1)\proj{k}U^\varepsilon(1)^\dag-\proj{N-k-1} \; \Vert
\end{eqnarray*}
reaches its sup over the allowed compact space since the state of a (sufficiently regular) dynamical system at a finite time depends continuously on system parameters (see e.g.~\cite[theorem 3.5]{Khalil-NonLinSystems}).
The proof for one system is in two steps: first we prove that the hypotheses of the adiabatic theorem with gap condition are verified, then we apply the theorem to compute the image at $s=1$ of initial projector $\proj{k}$ in adiabatic approximation.

\emph{Step 1:}
By hypothesis (a), we have $H(s) \in \mathcal{C}^2([0,1],\mathcal{H}_N)$ and therefore continuous over $[0,1]$. From \cite[section II.5.2]{Kato_1966}, it is then possible to find $N$ continuous functions $\lambda_0(s),\ldots,\lambda_{N-1}(s)$ such that $\lambda_0(s) \leq \ldots \leq \lambda_{N-1}(s)$ for all $s\in[0,1]$ are the eigenvalues of $H(s)$.
In terms of associated eigenspace projections, note that $\{P_{\lambda_k(s)} : k \in \NN_0^{N\minou 1} \} = \{\proj{k} : k\in \NN_0^{N\minou 1}\}$ every time $A(s)=0$, by unicity of the spectral decomposition of a non degenerate matrix. However, the pairwise correspondence depends on the value of $\omega(s)$. In particular, by hypotheses (b) and (c),
\begin{equation}\label{eq:Pk1}
P_{\lambda_k(0)} = \proj{k} \; \mbox{ and } \; P_{\lambda_k(1)} = \proj{N-k-1}
\end{equation}
for all $k$. For a given $s \in [0,1]$,
\newline $\bullet$ either $A(s)\neq 0$, then $H(s)$ has $N$ distinct eigenvalues according to Lemma \ref{Lemma:nondegeneracy};
\newline $\bullet$ or $A(s) = 0$, then $H(s)=H_R(\omega(s))$ and it must have $N$ distinct eigenvalues by hypothesis (c).
Hence,
\begin{equation}
\label{eq:eigenvalues}
\lambda_0(s)<..<\lambda_{N-1}(s) \mbox{ for all } s\in[0,1] \, .
\end{equation}
Then by continuity over the compact $[0,1]$, there exists $\delta>0$ such that $\lambda_{k}(s)+\delta<\lambda_{k+1}(s)$ for all $k \in \NN_0^{N\minou 2}$ and for all $s \in [0,1]$: each $\lambda_{k}(s)$ is at all times surrounded by a ``spectral gap'' of amplitude $\delta$ in which there is no other eigenvalue. We can therefore apply the adiabatic theorem with gap condition (see \cite[Theorem 2.2]{Teufel_2003}) to eigenvalue $\lambda_k(s)$, for any particular $k \in \NN_0^{N\minou 1}$, as is done in the following.

\emph{Step 2:} The adiabatic theorem ensures that $P_{\lambda_k(s)} \in \mathcal{C}^2([0,1],\mathcal{H}_N)$.
Define the ``adiabatic Hamiltonian''
\begin{equation}
\label{eq:Ha}
H_{a,k}(s)=H(s)- i\varepsilon P_{\lambda_k(s)} \frac{d}{ds}P_{\lambda_k(s)}-i \varepsilon P_{\lambda_k(s)}^\bot\frac{d}{ds}P_{\lambda_k(s)}^\bot
\end{equation}
where $P_{\lambda_k(s)}^\bot=I-P_{\lambda_k(s)}$, and the ``adiabatic propagator'' $U_{a,k}^\varepsilon$ which verifies, for all $s\in[0,1]$,
\begin{equation}
\label{eq:Ua}
i\varepsilon\, \frac{d}{ds}\, U_{a,k}^\varepsilon(s)=H_{a,k}(s)\, U_{a,k}^\varepsilon(s) \;\; \mbox{ with } U_{a,k}^\varepsilon(0)=I\; .
\end{equation}
One verifies that this construction ensures
\begin{equation}\label{eq:Haprop}
U^\varepsilon_{a,k}(s)P_{\lambda_k(0)}{U^\varepsilon_{a,k}(s)}^\dag=P_{\lambda_k(s)}
\end{equation}
for all $s\in[0,1]$. The adiabatic theorem states the existence of a constant $C_1>0$ such that
\begin{equation*}
\norm{U^\varepsilon(s)-U^\varepsilon_{a,k}(s)}\le C_1 \varepsilon\;\; \mbox{ for all } \; s\in[0,1]\; ,
\end{equation*}
in particular for $s=1$. This implies
\begin{eqnarray*}
	\lefteqn{\Vert U^\varepsilon(1)\proj{k}U^\varepsilon(1)^\dag - U^\varepsilon_{a,k}(1)\proj{k}U^\varepsilon_{a,k}(1)^\dag \Vert} \\
	& \;\; \leq & \Vert (U^\varepsilon(1)-U^\varepsilon_{a,k}(1))\proj{k} U^\varepsilon(1)^\dag \Vert\\ & & +\;\; \Vert U^\varepsilon_{a,k}(1)\proj{k} (U^\varepsilon(1)-U^\varepsilon_{a,k}(1))^\dag \Vert\\
	& \;\; \leq & \Vert U^\varepsilon(1)-U^\varepsilon_{a,k}(1) \Vert\;\; \Vert\, \proj{k}\, \Vert \;\; (\Vert U^\varepsilon_{a,k}(1) \Vert + \Vert U^\varepsilon(1) \Vert) \\
	& \;\; \leq & C_1 \varepsilon \;\cdot\; 1 \;\cdot\; 2\sqrt{N}
\end{eqnarray*}
since $\Vert U \Vert = \sqrt{\tr{U^\dag U}} = \sqrt{\tr{I}}$ for any unitary matrix $U$. Combining this with \eref{eq:Pk1},\eref{eq:Haprop} yields the result, where $C= 2C_1 \sqrt{N}$.
\hfill $\boldsymbol{\Box}$
\end{pf}


\subsection{Proof of Theorem \ref{thm:2} and corollary \ref{crl:1}}

We start by proving a Lemma about the behavior of time-dependent eigenvalues crossing each other.
\begin{lm}
\label{Lemma:eigenvalueswitching}
Assume that $H(s)$ as defined in \eref{eq:hamiltonians} depends analytically on the real parameter $s$ on an interval $\mathcal{I}\subset\RR$, with $\frac{d}{ds}\omega(s)>\gamma>0$ for all $s\in \mathcal{I}$. Suppose that $H_R(\omega(s))$ is non degenerate on $\mathcal{I}$ except for a simple degeneracy at $\bar s \in \mathcal{I}$, i.e.~$H_R(\omega(\bar s))$ has $N-1$ distinct eigenvalues and $H_R(\omega(s))$ has $N$ distinct eigenvalues for $s \in \mathcal{I}\setminus\{\bar s\}$. If $A(\bar s)=0$, then:
\begin{enumerate}
\item[(a)] There exist $N$
unique
functions $\lambda_0,\ldots,\lambda_{N-1}$ analytic over $\mathcal{I}$, with $\lambda_0(s)<\ldots<\lambda_{N-1}(s)$ for all $s<\bar s$, and such that $\{\lambda_0(s),\ldots,\lambda_{N-1}(s)\}$ are the eigenvalues of $H(s)$ for all $s\in \mathcal{I}$.
\item[(b)] Let $k$ be such that $\lambda_k(\bar s)=\lambda_{k+1}(\bar s)$. Then for all $s>\bar s$ we have
\begin{equation*}
\lambda_0(s)<\ldots<\lambda_{k+1}(s)<\lambda_k(s)<\ldots<\lambda_{N-1}(s) \; .
\end{equation*}
\end{enumerate}
\end{lm}

\begin{pf}[of Lemma \ref{Lemma:eigenvalueswitching}] Point (a) is a direct consequence of \cite[Theorem 6.1]{Kato_1966}. The order of the analytic eigenvalues is obviously preserved over time intervals where $H(s)$ is non degenerate; by Lemma \ref{Lemma:nondegeneracy}, these intervals are $\{ s < \bar s \}$ and $\{ s > \bar s \}$. The issue is what happens at $s=\bar s$. In the following, we show that $\lambda_k'(\bar s) \neq \lambda_{k+1}'(\bar s)$. Since the eigenvalues are analytic and $\lambda_k(\bar s)=\lambda_{k+1}(\bar s)$, a Taylor expansion then
yields the conclusion of (b).

We lead calculations similar to those of \cite[section XVI.II.8]{Messiah58}.
According to \cite[section II.6.2]{Kato_1966}, since $H$ is analytic over $\mathcal{I}$ and $H(s)\in\mathcal{H}_N$ for all $s\in\mathcal{I}$, there exist rank one orthogonal spectral projections $P_{\lambda_0(s)},\ldots,P_{\lambda_{N-1}(s)}$ which are analytic over $\mathcal{I}$. Computing the derivative of
\begin{equation}
\label{eq:proj}
H(s)P_{\lambda_k(s)}=\lambda_k(s)P_{\lambda_k(s)}
\end{equation}
with respect to $s$ at $s=\bar s$, we get
$$ H'(\bar s)P_{\lambda_k(\bar s)}+H(\bar s)P'_{\lambda_k(\bar s)}=\lambda_k'(\bar s)P_{\lambda_k(\bar s)}+\lambda_k(\bar s)P'_{\lambda_k(\bar s)}\; .$$
Multiplying the last equation by $(P_{\lambda_k(\bar s)}+P_{\lambda_{k+1}(\bar s)})$ from the left, using \eref{eq:proj} and the fact that $P_{\lambda_k}$ and $P_{\lambda_{k+1}}$ are two orthogonal projectors ($P_{\lambda_k}^2=P_{\lambda_k}$, $P_{\lambda_{k+1}}^2=P_{\lambda_{k+1}}$ and $P_{\lambda_k}\, P_{\lambda_{k+1}}=0$), we get $(P_{\lambda_k(\bar s)}+P_{\lambda_{k+1}(\bar s)})H'(\bar s)P_{\lambda_k(\bar s)}=\lambda_k'(\bar s)P_{\lambda_k(\bar s)}$. Noting that $P_{\lambda_k(\bar s)}=(P_{\lambda_k(\bar s)}+P_{\lambda_{k+1}(\bar s)})P_{\lambda_k(\bar s)}$, we get
\begin{eqnarray*}
(P_{\lambda_k(\bar s)}+P_{\lambda_{k+1}(\bar s)})H'(\bar s)(P_{\lambda_k(\bar s)}+P_{\lambda_{k+1}(\bar s)})P_{\lambda_k(\bar s)}\\
=\lambda_k'(\bar s)P_{\lambda_k(\bar s)}\; .
\end{eqnarray*}
The analog holds with $k$ and $k+1$ switched. This implies that $\{\lambda_k'(\bar s),\lambda_{k+1}'(\bar s)\}$ are the eigenvalues of the $2\times 2$ matrix obtained by restricting operator $H'(\bar s)$ to the column space of $(P_{\lambda_k(\bar s)}+P_{\lambda_{k+1}(\bar s)})$.
Since $A(\bar s)=0$ we have $H(\bar s)=H_R(\omega(\bar s))$. Denoting $\ket{m}$ and $\ket{n}$ the two eigenvectors of $H_R$ corresponding to eigenvalue $\lambda_k(\bar s)=\lambda_{k+1}(\bar s)$,
we have $P_{\lambda_k(\bar s)}+P_{\lambda_{k+1}(\bar s)}=\proj{m}+\proj{n}$. Defining
$$\left(H'(\bar s)\right)_{mn}=
\left(\begin{array}{cc}
\bra{m}H'(\bar s)\ket{m} & \bra{m}H'(\bar s)\ket{n}\\
\bra{n}H'(\bar s)\ket{m} & \bra{n}H'(\bar s)\ket{n}
\end{array}\right)$$
and computing
\begin{equation*}
H'(\bar s)=\omega'(\bar s) {\frac{d}{dv}H_R(v)}\mid_{v=\omega(\bar s)}+A'(\bar s)H_1 \, , \mbox{ we get}
\end{equation*}
\begin{equation}
\label{eq:dH/dsmn}
\left(H'(\bar s)\right)_{mn}=
\left(\begin{array}{cc}
-m\omega'(\bar s) & A'(\bar s)\mu_{mn}\\
A'(\bar s)\mu_{mn} & -n\omega'(\bar s)
\end{array}\right)
\end{equation}
where $\mu_{mn}=\bra{m}H_1\ket{n}$. Thus $\mu_{mn}=0$ if $|m-n|>1$ and $\mu_{mn}\ne 0$ if $|m-n|=1$. In both cases, since $\omega'(\bar s)\ne0$ and $m\ne n$, the matrix in \eref{eq:dH/dsmn} has $2$ real and distinct eigenvalues, corresponding to $\lambda_k'(\bar s) \neq \lambda_{k+1}'(\bar s)$.
\hfill $\boldsymbol{\Box}$
\end{pf}

\begin{pf}[of Theorem \ref{thm:2}] Taking $A(s) = 0$ at some points where $H_R$ is degenerate means that eigenvalues of $H(s)$ will not remain distinct at those points. We therefore use the adiabatic theorem without spectral gap condition, see \cite[corollary 2.5]{Teufel_2003}. Like for Theorem \ref{thm:1}, we prove the result for one system $\in \SSS$ and conclude the result for the sup. The proof is again in two steps. First we state how the adiabatic theorem can be applied; then we compute the image at $s=1$ of initial state $\proj{k}$ in adiabatic approximation.

\emph{Step 1:} Since $H$ is Hermitian, analytic over $[0,1]$ and simply degenerate at isolated points, we can apply Lemma \ref{Lemma:eigenvalueswitching}(a) repeatedly to conclude that there is a unique set of functions $\lambda_0,\ldots,\lambda_{N-1}$ analytic over $\mathcal{I}$, with $\lambda_0(0)<\ldots<\lambda_{N-1}(0)$,
and such that $\{\lambda_0(s),\ldots,\lambda_{N-1}(s)\}$ are the eigenvalues of $H(s)$ for all $s\in \mathcal{I}$. Moreover, according to \cite[section II.6.2]{Kato_1966}, there is a unique set of associated rank-one projectors $P_{\lambda_0(s)},\ldots,P_{\lambda_{N-1}(s)}$ which are analytic over $\mathcal{I}$.
In particular, given assumption (A3) and as $H(s)=H_R(\omega(s))$ for $s\in \{0,1\}$, we have $(\lambda_k(0), P_{\lambda_k(0)}) = (\lambda_k^R(0), \proj{k})$ for all $k$ and $\{(\lambda_0(1),P_{\lambda_0(1)}),\ldots,(\lambda_{N-1}(1),P_{\lambda_{N-1}(1)})\} = \{(\lambda_0^R(1),\proj{0}),\ldots,(\lambda_{N-1}^R(1),\proj{N-1})\}$. Note however that, unlike for Theorem \ref{thm:1}, the pairwise correspondence between elements of the latter sets is not obvious a priori, because here eigenvalues of $H(s)$ do not remain distinct on $[0,1]$.
A second difficulty is to assess how the system's state evolves when eigenvalues become degenerate. This second part is answered by the adiabatic theorem witout gap condition.
Introduce, as in Theorem \ref{thm:1}, the adiabatic Hamiltonian $H_{a,0}$ and adiabatic propagator $U^\varepsilon_{a,0}$, given by \eref{eq:Ha} and \eref{eq:Ua} respectively with $k=0$. Then by construction $U^\varepsilon_{a,0}(1)\proj{0}U^\varepsilon_{a,0}(1)^\dag = U^\varepsilon_{a,0}(1)P_{\lambda_0(0)} U^\varepsilon_{a,0}(1)^\dag = P_{\lambda_0(1)}$.
The adiabatic theorem states that $\exists C$ such that
\begin{equation}
\label{eq:adiabaticTheoremwithouthgap}
\Vert \, U^\varepsilon(s)\proj{k}U^\varepsilon(s)^\dag-U^\varepsilon_{a,0}(s)\proj{k}U^\varepsilon_{a,0}(s)^\dag \, \Vert \; \leq \;\; C \sqrt{\varepsilon}
\end{equation}
for all $\ket{k} \in \{\ket{0},\ldots,\ket{N-1}\}$. Thus the actual system adiabatically follows the analytic $P_{\lambda_k(s)}$, from $P_{\lambda_0(0)}=\proj{0}$ up to $P_{\lambda_0(1)}$ in particular.

\emph{Step 2: }We now compute $P_{\lambda_0(1)}$.
Define a small interval $I_{mn} = [\tau^o_{mn},\tau^f_{mn}] \subset [0,1]$ around each point $s(m,n)$ such that all $I_{mn}$ are disjoint. If $A(s(m,n))\neq 0$, then as shown in Theorem \ref{thm:1}, $H(s)$ is non degenerate for all $s \in I_{mn}$, such that for any $j,k$ with $\lambda_j(\tau^o_{mn}) < \lambda_k(\tau^o_{mn})$ we have $\lambda_j(\tau^f_{mn}) < \lambda_k(\tau^f_{mn})$. On the other hand, if we take $A(s(m,n))=0$, then two eigenvalues intersect at $s=s(m,n)$ and the analytic branches cross so that their order changes as stated in Lemma \ref{Lemma:eigenvalueswitching}(b). To avoid separate treatment of limit cases, we define $s_0=0$ and $s_N=1$. Now by construction:
\begin{itemize}
\item $\lambda_j(0) = \lambda^R_j(0)$ for all $j \in \NN_0^{N\minou 1}$.
\item For $k \in \NN_1^{N}$, $\lambda^R_0(s)$ is the $k^{\mbox{th}}$ smallest eigenvalue of $H_R(\omega(s))$ when $s \in (s_{k-1},s_k)$.
\item As long as $A(s_0)=\ldots=A(s_{k-1})=0$, that is for $k\leq N-p$, $\lambda_0(s)$ follows the same crossings as $\lambda^R_0(s)$; therefore it is the $k^{\mbox{th}}$ smallest eigenvalue of $H(s)$ when $s \in (s_{k-1},s_k)$.
\item For $s>s_{N-p-1}$, we have $A(s) \neq 0$ so the $\lambda_k(s)$ keep the same order, i.e.~$\lambda_0(s)$ remains the $(N-p)^{\mbox{th}}$ smallest eigenvalue of $H(s)$.
\item In particular for $s=1$, from \eref{eq:ineqeigen} we identify $\lambda_0(1) = \lambda^R_{N-(N-p)}(1) = \lambda^R_p(1)$, such that $P_{\lambda_0(1)} = \proj{p}$ by uniqueness of the spectral decomposition. \hfill $\boldsymbol{\Box}$
\end{itemize}
\end{pf}

\begin{rmk}
To apply the adiabatic Theorem \cite[corollary 2.5]{Teufel_2003}, it is sufficient to have $P_{\lambda_0(s)} \in \mathcal{C}^2([0,1],\mathcal{H}_N)$. However, a condition like $H(s) \in \mathcal{C}^2([0,1],\mathcal{H}_N)$ does not ensure the existence of $P_{\lambda_0(s)},\ldots,P_{\lambda_{N-1}(s)} \in \mathcal{C}^2([0,1],\mathcal{H}_N)$, see \cite[example 5.3]{Kato_1966}. It is only for analytic $H(s)$ that we can guarantee analytic $P_{\lambda_0(s)}$, which then in particular belongs to $\mathcal{C}^2([0,1],\mathcal{H}_N)$.
\end{rmk}

\begin{pf}[of corollary \ref{crl:1}]
The arguments are the same as in the proof of Theorem \ref{thm:2}. We concentrate on tracking the analytic eigenvalue branches $\lambda_0(s),\ldots,\lambda_{N-1}(s)$ of $H(s)$ to establish their pairwise correspondence with eigenvalues $\lambda^R_0(s),\ldots,\lambda^R_{N-1}(s)$ of $H_R(\omega(s))$ at $s=1$.
We prove the result for $p < N-l-1$; the case $p > N-l-1$ is treated similarly, while $p = N-l-1$, implying $\mathcal{I}^\omega_{lp} = \emptyset$, is the case covered by Theorem \ref{thm:1}. Denote $s_1<\ldots<s_{N-l-p-1}$ the elements of $\mathcal{I}^\omega_{lp}$, and $s_0=0$, $s_{N-l-p}=1$.

The algorithm constructs $\mathcal{I}^\omega_{lp}$ such that the $(l+d)^{\mbox{th}}$ and $(l+d+1)^{\mbox{th}}$ smallest eigenvalues of $H_R(\omega(s))$ become equal at $s_d$, for each $d\in \NN_1^{N\minou l \minou p \minou 1}$. Taking $A(s_d) = 0$ implies $H(s_d)=H_R(\omega(s_d))$ so the same eigenvalue equalities hold for $H(s)$ at $s=s_d$. Moreover from point (c) and Lemma \ref{Lemma:nondegeneracy} all eigenvalues of $H(s)$ remain distinct for $s \not\in \mathcal{I}^\omega_{lp}$.
Therefore the analytic eigenvalue branch $\lambda_l(s)$, starting with $\lambda_l(0)=\lambda^R_l(0)$, exactly evolves through crossings at $s_1,\ldots,s_{N-l-p-1}$ such that it is the $(l+d+1)^{\mbox{th}}$ smallest eigenvalue of $H(s)$ for $s \in (s_d,s_{d+1})$. In particular, $\lambda_l(1)$ is the $(N-p)^{\mbox{th}}$ smallest eigenvalue of $H(1)=H_R(\omega(1))$, which from (A2) means $\lambda_l(1)=\lambda^R_{p}(1)$ such that $P_{\lambda_l(1)} = \proj{p}$.
\hfill $\boldsymbol{\Box}$
\end{pf}


\section{Summary and discussion}\label{sec:conclusion}

This paper shows how adiabatic passage can be applied to a quantum ladder system to achieve permutations of populations on the ladder levels with a single laser pulse. We explicitly propose control inputs whose precise functional dependence on time is not important as long as they satisfy a few key features, most notably annihilation or not at specific times. This makes our strategy robust against multiplicative input disturbances. Another important advantage of our adiabatic strategy is its ability to simultaneously control an ensemble of systems with different dipole moment values.

Theorems in the present paper provide a proof of concept in idealized situations. Several practical issues deserve a more quantitative investigation in future work. Probably the most important aspect is to characterize precision of the adiabatic approximation as a function of $\varepsilon$. Indeed, for small $\varepsilon$ the actual control time $t=\frac{s}{\varepsilon}$ gets long; this further implies that, at constant power $A^2(s)$, the energy given to the system gets large. Beyond performance requirements, this also invalidates our model at infinitesimal $\varepsilon$ (e.g.~regarding finite lifetime of the levels).
Although orders of magnitude are given for the adiabatic limit, variations in the proportionality constant can lead to significant discrepancies. Investigating them, as well as ``optimal paths'' minimizing non-adiabatic losses \cite{Guerin-Thomas-Jauslin-PRA_2002}, could yield guidelines for choosing amongst several possible ``eigenvalue crossing designs''.
Both precision of adiabatic approximation and modeling assumptions (e.g.~RWA) also limit the range of ``ensemble'' properties in practice.

It may appear surprising at first sight that two different evolutions are selected just by taking $A(\overline{s})=0$ or $A(\overline{s})\neq 0$ at a precise instant $\overline{s}$. The elucidation is that this dichotomy only holds at the limit $\varepsilon \rightarrow 0^+$. For a given $\varepsilon$, the larger $\vert A \vert$ in the neighborhood of $s=\overline{s}$, the more the evolution differentiates from the $A(\overline{s})=0$ case. Nevertheless, for small $\varepsilon$, the relevant neighborhood around $\overline{s}$ for selecting population transfer or not indeed gets small (from there experimentalists' denomination ``rapid adiabatic passage''). Our scheme might therefore allow selective population permutation as a function of $\{\Delta_0,\ldots,\Delta_{N-1}\}$ on an ensemble of systems, in a scheme resembling resonance selection. Take $A(\overline{s})=0$ for $\overline{s} \in \mathcal{I}^\omega$ of a nominal system. If a system has detunings very close to nominal, then two of its $\lambda^R_k(s)$ cross at a point $\widetilde{s}$ close to $\overline{s}$, where $A(\widetilde{s}) \approx 0$, such that for moderate $\varepsilon$ its final state will be close to the adiabatic result of the nominal system with $A(\overline{s})=0$. If a system has detunings more different from nominal, then all its crossings of $\lambda^R_k(s)$ occur at points where $A$ significantly differs from zero, and with moderate $\varepsilon$ its final state will be closer to the adiabatic result of the nominal system with $A(\overline{s})\neq 0$. A quantitative statement of this idea requires further investigation.


\begin{ack}
The authors thank M.~Lemaire for useful discussions. ZL acknowledges support from ANR, Projet Jeunes Chercheurs EPOQ2 number ANR-09-JCJC-0070.  AS is an FNRS postdoctoral researcher. This paper presents research results of the Belgian Network DYSCO (Dynamical Systems, Control, and Optimization), funded by the Interuniversity Attraction Poles Programme, initiated by the Belgian State, Science Policy Office. The scientific responsibility rests with its authors.  PR acknowledges support from ANR (CQUID).
\end{ack}


\section*{References}
\bibliographystyle{unsrt}
\bibliography{Adiabib}

\end{document}